\documentclass{aa}

\usepackage{natbib}
\bibpunct{(}{)}{;}{a}{}{,}

\usepackage{graphicx}
\usepackage{txfonts}
\newcommand{\va}{v_{\mathrm{A}}}
\newcommand{\vaj}{v_{\mathrm{Aj}}}
\newcommand{\vae}{v_{\mathrm{Ae}}}
\newcommand{\vai}{v_{\mathrm{Ai}}}

\newcommand{\td}{\tau_{\mathrm{D}}}
\newcommand{\tdp}{\tau_{\mathrm{D}} / P}

\newcommand{\re}{\rho_{\rm e}}
\newcommand{\rj}{\rho_{\rm j}}
\newcommand{\ri}{\rho_{\rm i}}

\newcommand{\raj}{r_{\rm A,j}}

\newcommand{\xii}{\mbox{\boldmath{$\xi$}}}

\begin{document}

\title{Damped transverse oscillations of interacting coronal loops}

\titlerunning{Damped collective loop oscillations}

   \author{Roberto Soler\inst{1,2} \& Manuel Luna\inst{3,4}}

   \institute{Departament de F\'isica, Universitat de les Illes Balears, E-07122 Palma de Mallorca, Spain.    
   \and
  Institut d'Aplicacions Computacionals de Codi Comunitari (IAC$^3$), Universitat de les Illes Balears, E-07122 Palma de Mallorca, Spain. 
              \and
Instituto de Astrof\'isica de Canarias, E-38200 La Laguna, Tenerife, Spain.   
\and
Departamento de Astrof\'isica, Universidad de La Laguna, E-38206 La Laguna, Tenerife, Spain. \\
\email{roberto.soler@uib.es, mluna@iac.es} }

   \date{Received XXX; accepted XXX}

 
  \abstract
   {Damped transverse oscillations of magnetic loops are routinely observed in the solar corona. This phenomenon is  interpreted as standing kink magnetohydrodynamic  waves, which are damped by resonant absorption owing to plasma inhomogeneity across the magnetic field. The  periods and damping times of these oscillations can be used to probe the physical conditions of the coronal medium. Some observations suggest that  interaction between neighboring oscillating loops in an active region may be important and can modify the properties of the  oscillations compared to those of an isolated loop. Here we theoretically investigate resonantly damped transverse oscillations of interacting non-uniform coronal loops. We provide a semi-analytic method, based on the T-matrix theory of scattering, to compute the frequencies and damping rates of  collective oscillations of an arbitrary configuration of parallel cylindrical  loops. The effect of resonant damping is included in the T-matrix scheme in the thin boundary approximation.  Analytic and numerical results in the specific case of two interacting loops are given as an application.}


   \maketitle
%

\section{INTRODUCTION}

Transverse oscillations of magnetic loops in the solar corona are under intense research since the first observational reports by the {\em Transition Region And Coronal Explorer} (TRACE)  \citep[see, e.g.,][]{1999Sci...285..862N,1999ApJ...520..880A}. Large-amplitude  coronal loop  oscillations are usually excited after energetic events as, e.g., solar flares, coronal mass ejections, or low coronal eruptions \citep[see][]{2015A&A...577A...4Z}. Based on magnetohydrodynamic (MHD) wave theory \citep[e.g.,][]{2005LRSP....2....3N},  transverse loop oscillations have been interpreted as standing kink  MHD  waves. Kink MHD modes are nearly incompressible waves,  mainly driven by magnetic tension, and  responsible for global transverse motions of the flux tube \citep[see, e.g.,][]{1983SoPh...88..179E,2009A&A...503..213G,2012ApJ...753..111G}.  A relevant feature of  large-amplitude loop oscillations is that they are strongly damped.   It has been shown that resonant absorption, caused by plasma inhomogeneity in the direction perpendicular to the magnetic field, is an efficient damping mechanism of kink MHD waves in coronal loops \citep[see, e.g.,][]{2002ApJ...577..475R,2002A&A...394L..39G}.  Due to resonant absorption, the energy from the global  kink motion of the flux tube is transferred to small-scale, unresolved rotational motions around the nonuniform boundary of the tube \citep[see, e.g.,][]{2006ApJ...642..533T,2014ApJ...788....9G,2015ApJ...803...43S}.  As a result of this process, the  global kink oscillation of the coronal loop is quickly damped in time. The interested reader is referred to \citet{2011SSRv..158..289G}, where the theory and applications of resonant waves in the solar atmosphere are reviewed.

Observations often show that neighboring oscillating  loops in an active region interact with each other and exhibit collective behaviour \citep[e.g.,][]{2000ApJ...537L..69S,2004SoPh..223...77V,2002SoPh..206...69S,2013ApJ...774..104W}. Interaction between loops can modify the properties of their transverse oscillations compared to those of the classic kink mode of an isolated loop. Therefore, advanced models describing coronal loop oscillations should take into account interactions within loop systems. A number of works have studied collective transverse oscillations  in Cartesian geometry \citep[e.g.,][]{2005A&A...440.1167D,2006SoPh..236..111D,2006A&A...457.1071L,2007A&A...466.1145A,2008ApJ...674.1179A}. In cylindrical geometry, \citet{2008ApJ...676..717L} numerically investigated transverse oscillations of two cylindrical loops, and \citet{2009ApJ...694..502O} performed numerical simulations in the case of four interacting loops. Concerning analytical works in cylindrical geometry, \citet{2009ApJ...692.1582L,2010ApJ...716.1371L}  used the  T-matrix theory of scattering \citep[see, e.g.,][]{1952ASAJ...24...42T,1969ASAJ...45.1417W,1987ApJ...318..888B,1994ApJ...436..372K} to investigate transverse oscillations of two and three loops \citep{2009ApJ...692.1582L} and of bundles of many loops \citep{2010ApJ...716.1371L}  in the $\beta=0$ approximation, where $\beta$ refers to the ratio of the gas pressure to the magnetic pressure.   \citet{2009ApJ...693.1601S} later extended the method of  \citet{2009ApJ...692.1582L,2010ApJ...716.1371L}  by incorporating  gas pressure and longitudinal flows and studied collective oscillations of flowing prominence threads. These works showed that loop interaction affects the properties of their oscillations.  \citet{2008ApJ...676..717L,2009ApJ...692.1582L} obtained that a system of two loops of arbitrary radii supports four kink-like collective modes. The shift of the collective mode frequencies with respect to the individual kink frequencies of the loops is significant when the distance between loops is small (of the order of the loop radius) and when loops have similar densities. Conversely, the oscillating loops show little interaction when they are far from each other and when their densities are substantially different. On the other hand, \citet{2008A&A...485..849V} and \citet{2010A&A...515A..33R} used a different method based on bycilindrical coordinates to study transverse oscillations of two pressure-less loops in the thin tube (TT) approximation. Of the four kink-like collective modes obtained by \citet{2009ApJ...692.1582L} in the T-matrix theory, only two different modes remain in the TT approximation considered by \citet{2008A&A...485..849V} and \citet{2010A&A...515A..33R}.

Concerning the damping of the oscillations, \citet{2007A&A...466.1145A,2008ApJ...674.1179A} investigated resonantly damped oscillations of two slabs, while \citet{2008ApJ...679.1611T} numerically studied the resonant damping of transverse oscillations of a multi-stranded loop. Those works showed that the process of resonant damping is not compromised by the irregular geometry of a realistic loop model and still produces the efficient attenuation of  global transverse oscillations.  The damping of transverse oscillations  of two cylindrical loops was analytically  investigated by \citet{2011A&A...525A...4R} and \citet{2014A&A...562A..38G}, who considered the TT approximation and used  bycilindrical coordinates. Results obtained with bycilindrical coordinates should be treated with caution  when the distance between loops is small. Geometrical effects intrinsically associated to the bycilindrical coordinates  may produce unphysical results. Our purpose is to use the T-matrix method of \citet{2009ApJ...692.1582L,2010ApJ...716.1371L} to investigate resonantly damped oscillations of bundles of loops.  The present paper is partially based on unpublished results included in \citet{2010PhDT.........3S}\footnote{The full text of \citet{2010PhDT.........3S} is available at \url{http://www.uib.es/depart/dfs/Solar/thesis_robertosoler.pdf}}. The effect of resonant absortion in the Alfv\'en continuum  is incorporated to the T-matrix scheme  by using the method that combines the jump conditions of the perturbations at the resonance position with the so-called thin boundary (TB) approximation \citep[see, e.g.,][]{1991SoPh..133..227S,1992SoPh..138..233G}. A similar method has  previously been used by \citet{1995SoPh..161..251K} to investigate absorption of acoustic waves. We provide a general analytic theory, which is valid for bundles of many transversely nonuniform parallel loops of arbitrary radii. Specific results in the case of two loops are obtained  and   compared to those given in \citet{2011A&A...525A...4R} and \citet{2014A&A...562A..38G}. Our results are also compared to those of \citet{2007A&A...466.1145A,2008ApJ...674.1179A} obtained in Cartesian geometry.

This paper is organized as follows. Section~\ref{sec:model} contains the description of the equilibrium configuration and the basic governing equations. The general analytic T-matrix theory of scattering to compute the frequencies and damping rates of collective loop oscillations is given in Section~\ref{sec:theory}. Later, the specific case of damped oscillations of two loops is discussed both analytically and numerically in Section~\ref{sec:twoloops}. Finally, some concluding remarks are given in Section~\ref{sec:conclusion}.

\section{MODEL AND GOVERNING EQUATIONS}
\label{sec:model}

\begin{figure}[!t]
\centering
\includegraphics[width=.95\columnwidth]{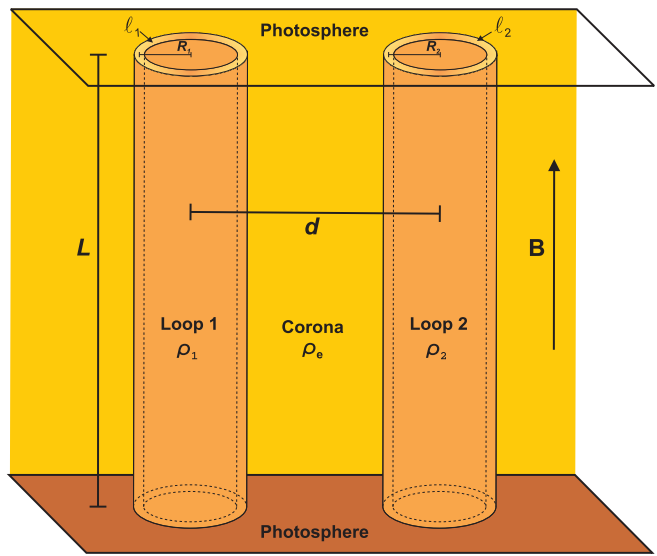}
\caption{Sketch of the equilibrium configuration in the specific case of two transversely non-uniform coronal loops. \label{fig:model}}
\end{figure}

Our equilibrium configuration is composed of $N$ straight and parallel magnetic cylinders of length $L$ embedded in a uniform coronal plasma. The ends of the magnetic tubes are fixed at two rigid walls representing the solar photosphere. We set the $z$-direction to be along the axes of the tubes. The magnetic field is straight along the $z$-direction, namely ${\bf B} = B \hat{e}_z$, where $B$ is a constant everywhere.  We use subscripts `i' and `e' to refer to, in general, the internal region of the tubes and the external plasma, respectively. The subscript or superscript `j' is used to refer to a particular loop.  We denote by $R_{\rm j}$ the radius of the jth tube. The distance between the centers of the jth and j'th loops is $d_{\rm jj'}$.  We denote by $\rj$ the internal density of the jth tube, while $\re$ denotes the external density, i.e., the density of the coronal environment. In our model $\rj$ and $\re$ are constants. There is a transversely nonuniform transitional layer surrounding each magnetic tube in which the density continuously varies from the internal density, $\rj$, to the external density, $\re$. The thicknesses of the non-uniform boundary layer of the jth cylinter is $l_{\rm j}$.   A sketch of the equilibrium configuration in the case of two magnetic tubes ($N=2$) is given in Figure~\ref{fig:model}.

We adopt the $\beta=0$ approximation, where $\beta$ refers to the ratio of the thermal pressure to the magnetic pressure. This is an appropriate approximation to investigate transverse waves in the solar corona. In the $\beta=0$ approximation, the ideal MHD equations governing linear perturbations superimposed on the static equilibrium state are
\begin{eqnarray}
\rho \frac{\partial^2 \xii}{\partial t^2} &=& \frac{1}{\mu} \left(\nabla \times {\bf b}\right) \times {\bf B}, \label{eq:basic1}\\
{\bf b}&=& \nabla \times \left( \xii \times {\bf B} \right), \label{eq:basic2}
\end{eqnarray}
where $\xii$ is the plasma Lagrangian displacement, ${\bf b}$ is the magnetic field Eulerian perturbation, $\rho$ is the density, and $\mu$ is the magnetic permittivity. 

We assume the temporal dependence of  perturbations as  $\exp \left( -i \omega t \right)$, where $\omega$ is the oscillation frequency. In the case of transversely nonuniform tubes, the global transverse oscillations are quasi-modes whose frequency is complex owing to damping by resonant absorption, i.e., $\omega =  \omega_{\rm R} + i \omega_{\rm I}$, where $\omega_{\rm R}$ and $\omega_{\rm I}$ are  the real and imaginary parts of the frequency, respectively. The real part of $\omega$ is related to the period and the imaginary part corresponds to the damping rate of the oscillations. We consider that the oscillating flux tubes are line-tied at the photosphere, which  acts as a perfectly reflecting wall in this model owing to its large density compared to the coronal density. Hence, we assume  perturbations to be proportional to $\exp \left( i k_z z \right)$, with $k_z$ the longitudinal wavenumber. For standing oscillations, $k_z$  given by
\begin{equation}
k_z = \frac{n \pi}{L}, \qquad \textrm{with} \quad n = 1, 2, \dots
\end{equation}
We shall restict ourselves to the fundamental mode of oscillation, so we take $n=1$. 

In the regions with constant density, Equations~(\ref{eq:basic1}) and (\ref{eq:basic2}) can be reduced to following equation,
\begin{equation}
\nabla^2_\perp P'  + k_\perp^2 P' = 0, \label{eq:helm}
\end{equation}
where $P' = {\bf B} \cdot {\bf b} / \mu$ is the total pressure Eulerian pertubation and the subscript $\perp$ refers to the direction perpendicular to the magnetic field. Thus, $\nabla^2_\perp$ denotes the perpendicular part of the $\nabla^2$ operator. In turn, the quantity $k_\perp$ plays the role of the perpendicular wavenumber and is defined as
\begin{equation}
k_\perp^2 = \frac{\omega^2 - \omega_{\rm A}^2}{\va^2},
\end{equation}
where $\omega_{\rm A}^2 = k_z^2 \va^2$ is the square of the Alfv\'en frequency and $\va^2 = B^2 / \mu \rho$ is the square of the Alfv\'en velocity. We stress that Equation~(\ref{eq:helm}) is only valid in the regions with constant density, so it does not apply within the nonuniform boundaries of the loops.

\section{T-MATRIX THEORY OF SCATTERING}
\label{sec:theory}

Equation~(\ref{eq:helm}) is the two-dimensional Helmholtz Equation. To solve Equation~(\ref{eq:helm}) we use the scattering theory in its T-matrix formalism \citep[see, e.g.,][]{1969ASAJ...45.1417W}. In the solar context, the T-matrix theory has previously been  used to investigate the scattering and absorption properties of bundles of magnetic flux tubes \citep[e.g.][among others]{1985ApJ...298..867B,1989ApJ...342..545B,1994ApJ...436..372K,1995SoPh..161..251K}. \citet{2009ApJ...692.1582L,2010ApJ...716.1371L} used of this technique to compute the eigenmodes of systems of magnetic tubes. Because of the inhomogeneity of the tubes in the transverse direction, the  modes with frequencies between the internal Alfv\'en frequencies of the loops and the external Alfv\'en frequency  are resonant in the Alfv\'en continuum. As a result, the oscillations are damped by resonant absorption. The effect of resonant absorption was not considered by \citet{2009ApJ...692.1582L,2010ApJ...716.1371L}. Here we extend their theory to consider resonant damping.

\subsection{Solutions in the Internal and External Plasmas}

 We use local polar coordinates associated to the jth loop. We denote by  $r_{\rm j}$ and $\varphi_{\rm j}$ the radial and azimuthal coordinates, respectively, of the coordinate system whose origin is located at the center of the jth tube. We can define an equivalent coordinate system in each tube. In this local coordinate system, the solution to Equation~(\ref{eq:helm}) in the internal region of the jth tube can be expressed as
\begin{equation}
P'^{\rm j}_{\rm i}= \sum_{m = - \infty}^\infty A_m^{\rm j} J_m \left( k_{\perp \rm j} r_{\rm j} \right) \exp \left( i m \varphi_{\rm j} \right), \label{eq:ptin}
\end{equation}
where $m$ is the azimuthal wavenumber, $J_m$ is the usual Bessel function of the first kind of order $m$, and $A_m^{\rm j}$ are constants. Unlike the case of isolated tubes \citep[see, e.g.,][]{1983SoPh...88..179E}, the solution is not entirely described by a single value of $m$. Because of interaction between  tubes, the values of $m$ are coupled. For transverse, kink-like oscillations  the dominant terms in the expansion are those with $m = \pm 1$, but the contribution from other $m$'s is not negligible unless the tubes are  far from each other \citep{2009ApJ...692.1582L}.

The solution to Equation~(\ref{eq:helm}) in the external region is written using the principle of superposition, which is applicable to linear waves.  The total external solution is computed by adding the net contributions of all flux tubes, namely
\begin{equation}
P'_{\rm e} = \sum_{\rm j} P'^{\rm j}_{\rm e},\label{eq:ptex}
\end{equation}
where $P'^{\rm j}_{\rm e}$ is the net contribution  of the jth tube to the  external solution. The key idea behind the scattering theory is that the solution of Equation~(\ref{eq:helm}) in the external plasma can be decomposed into several {\em fields} with different physical meanings, namely the {\em  total}, {\em exciting}, and {\em scattered fields}. Here we give an overview of the method. Interested readers are referred to \citet{2009ApJ...692.1582L,2010ApJ...716.1371L} for extensive explanations.

In the external plasma, the {\em total  field} associated to the jth cylinder  can be  expressed as
\begin{equation}
 P'^{\rm j}_{\rm total} =  \sum_{m = -\infty}^{\infty}  \left[  \alpha_{1m}^{\rm j} H_m^{(1)} \left(  k_{\perp \rm e} r_{\rm j}  \right)+\alpha_{2m}^{\rm j} H_m^{(2)} \left(  k_{\perp \rm e} r_{\rm j}  \right) \right] \exp \left( i m \varphi_{\rm j} \right), \label{eq:extfieldgenj} 
\end{equation}
where $H_m^{(1)}$ and $H_m^{(2)}$ are the usual Hankel functions of the first and second kind, respectively, and $\alpha_{1m}^{\rm j}$ and $\alpha_{2m}^{\rm j}$ constants.  The first term on the right-hand side of Equation~(\ref{eq:extfieldgenj}) represents outgoing waves from the jth tube, and the second term represents incoming waves toward the jth tube. Importantly, we note that 
\begin{equation}
   P'_{\rm e} \neq \sum_{\rm j} P'^{\rm j}_{\rm total}. 
\end{equation}
The reason for this inequality is that the outgoing wave associated to a particular tube contributes as an incoming wave for all the other tubes. In other words, $\sum_{\rm j} P'^{\rm j}_{\rm total}$ is not the net external solution. To overcome this problem,  the {\em total  field} associated to the jth cylinder,  $ P'^{\rm j}_{\rm total}$, is decomposed into a {\em scattered field}, $P'^{\rm j}_{\rm scat}$, and a {\em exciting field}, $P'^{\rm j}_{\rm excit}$. Conceptually, the {\em scattered field} represents the actual contribution of the various tubes to the net external solution, whereas the {\em exciting field} can be understood as the cross-talk mechanism responsible for  interaction between  flux tubes \citep[see details in, e.g.,][]{1989ApJ...342..545B}. 

The full net  external solution, $P'_{\rm e}$, is defined so that it corresponds to the sum of  the {\em scattered fields} associated to all tubes, namely
\begin{equation}
P'_{\rm e}= \sum_{\rm j}  P'^{\rm j}_{\rm scat}. \label{eq:extfindgen}
\end{equation}
Conversely,  the {\em exciting field} associated to the jth tube is defined as the difference between the full net contribution  and the scattered field of the jth tube,  namely
\begin{equation}
 P'^{\rm j}_{\rm excit} = P'_{\rm e}  - P'^{\rm j}_{\rm scat} = \sum_{\rm n \ne j} P'^{\rm n}_{\rm scat}. \label{eq:relextsct}
\end{equation}
\citet{1969ASAJ...45.1417W} introduced the T-matrix operator, ${\bf T}^{\rm j}$, which linearly relates the  scattered and exciting fields as
\begin{equation}
P'^{\rm j}_{\rm scat} =  {\bf T}^{\rm j}  P'^{\rm j}_{\rm excit}.
\end{equation}
\citet{1987ApJ...318..888B} showed that for cylindrical scatterers the T-matrix is diagonal, and \citet{1994ApJ...436..372K} gave an expression of its elements, namely
 \begin{equation}
  T_{mm}^{\rm j}  = \frac{1}{2} \left( 1 - \frac{\alpha_{1m}^{\rm j}}{ \alpha_{2m}^{\rm j}} \right), \label{eq:tmatrel}
 \end{equation}
where $\alpha_{1m}^{\rm j}$ and $\alpha_{2m}^{\rm j}$ are the same constants the appear in Equation~(\ref{eq:extfieldgenj}). We can use Equation~(\ref{eq:tmatrel}) to eliminate $\alpha_{1m}^{\rm j}$ and write all the expressions in terms of $\alpha_{2m}^{\rm j}$ alone. With the help of these last formulae, and after some algebraic manipulations using well-known properties of the Bessel functions, we can rewrite Equation~(\ref{eq:extfieldgenj}) as
\begin{equation}
  P'^{\rm j}_{\rm total}  = \sum_{m = -\infty}^{\infty}  2 \alpha_{2m}^{\rm j}\left[    J_m \left( k_{\perp \rm e} r_{\rm j}  \right) -  T_{mm}^{\rm j} H_m^{(1)} \left( k_{\perp \rm e} r_{\rm j}   \right) \right] \exp \left( i m \varphi_{\rm j} \right),
\end{equation}
from where it is straightforward to identify both  exciting and scattered fields, namely
\begin{eqnarray}
  P'^{\rm j}_{\rm excit} &=& \sum_{m = -\infty}^{\infty}  2 \alpha_{2m}^{\rm j}  J_m \left( k_{\perp \rm e} r_{\rm j}  \right)\exp \left( i m \varphi_{\rm j} \right), \\
P'^{\rm j}_{\rm scat} &=& - \sum_{m = -\infty}^{\infty}  2 \alpha_{2m}^{\rm j}  T_{mm}^{\rm j} H_m^{(1)} \left( k_{\perp \rm e} r_{\rm j}   \right)  \exp \left( i m \varphi_{\rm j} \right).
\end{eqnarray}
Finally, we use the expression of $P'^{\rm j}_{\rm scat}$ into Equation~(\ref{eq:extfindgen}) to arrive at the total net solution in the external plasma, namely
\begin{equation}
 P'_{\rm  e}  =  - \sum_{\rm j}\sum_{m=-\infty}^{\infty} 2 \alpha_{2m}^{\rm j} T_{mm}^{\rm j} H_m^{(1)} \left( k_{\perp \rm e} r_{\rm j} \right) \exp \left( i m \varphi_{\rm j} \right). \label{eq:ext}
\end{equation}

Equations~(\ref{eq:ptin}) and (\ref{eq:ext}) formally describe the total pressure perturbation in the interior and in the exterior of the tubes, respectively. However, we recall that these expressions do not apply in the nonuniform boundary layers. The T-matrix elements, $T_{mm}^{\rm j}$, contain the information about how the solutions are connected across the non-uniform boundaries of the tubes.

\subsection{T-matrix Elements in the Thin Boundary Approximation}

At this stage we incorporate the effect of the nonuniform boundary layers.  In the nonuniform layer of the jth tube the global wave modes are resonant in the Alfv\'en continuum at the resonant position, $r_{\rm j} =\raj$, where the global oscillation frequency matches the local Alfv\'en frequency. The resonant position, $\raj$, is defined through the resonant condition $\omega^2 = k^2_{z} \va^2 (\raj)$.  
We use the TB approximation and  restrict ourselves to $l_{\rm j}/ R_{\rm j} \ll 1$.  The TB approximation assumes that the jump of the perturbations across the resonant layer is the same as their jump across the whole nonuniform layer. Thus, the connection formulae of the wave perturbations across the resonance are used as jump conditions for the total pressure and the Lagrangian displacement at the boundaries of the tubes. This method and its applications have been reviewed by \citet{2011SSRv..158..289G}. The TB approximation within the formalism of the T-matrix theory has previously been  used by \citet{1994ApJ...436..372K} and \citet{1995SoPh..161..251K}. 

General expressions of the connection formulae for the perturbations across the resonant layer can be found in, e.g., \citet{1991SoPh..133..227S}.   In local coordinates, the connection formulae for the total pressure Eulerian pertubation, $P'$, and the  radial component of the Lagrangian displacement, $\xi_{r}$, at $r_{\rm j} = \raj$ are
\begin{equation}
\left[ P' \right] = 0, \qquad \left[ \xi_{r} \right] = - i \pi \frac{m^2 /\raj^2}{\left| \rho  \Delta_{\rm A} \right|_{\rm j}} P',  \label{eq:boundary}
\end{equation}
where $[X] = X_{\rm e} - X_{\rm j}$ denotes the jump of the quantity $X$ across the resonant layer and $\left| \rho  \Delta_{\rm A} \right|_{\rm j}$ is defined as
\begin{equation}
\left| \rho  \Delta_{\rm A} \right|_{\rm j} = \rho ( \raj ) \left| \frac{\rm d}{{\rm d}r_{\rm j}} \left( \omega^2 - k^2_{z} \va^2 \right) \right|_{\raj} = \omega^2 \left| \frac{{\rm d} \rho}{{\rm d}r_{\rm j}} \right|_{\raj},
\end{equation}
where we have used the resonant condition $\omega^2 = k^2_{z} \va^2 (\raj)$. After imposing the jump conditions given in Equation~(\ref{eq:boundary}), we obtain both the T-matrix elements and the dispersion relation of the collective modes. 

On the one hand, the T-matrix elements  are
\begin{eqnarray}
T_{mm}^{\rm j} &=&  \frac{\frac{k_{\perp \rm e}}{\re \left( \omega^2 - k_z^2 \vae^2 \right)}  \frac{J_m' \left( k_{\perp \rm e} R_{\rm j} \right)}{J_m \left( k_{\perp \rm e} R_{\rm j} \right)} - \frac{k_{\perp \rm j}}{\rj \left( \omega^2 - k_z^2 \vaj^2 \right)}  \frac{J_m' \left( k_{\perp \rm j} R_{\rm j} \right)}{J_m \left( k_{\perp \rm j} R_{\rm j} \right)} + i \pi \frac{m^2 / r_{\rm A j}^2}{\left| \rho  \Delta_{\rm A} \right|}}{\frac{k_{\perp \rm e}}{\re \left( \omega^2 - k_z^2 \vae^2 \right)}  \frac{{H^{(1)}_m}' \left( k_{\perp \rm e} R_{\rm j} \right)}{H^{(1)}_m \left( k_{\perp \rm e} R_{\rm j} \right)} - \frac{k_{\perp \rm j}}{\rj \left( \omega^2 - k_z^2 \vaj^2 \right)}  \frac{J_m' \left( k_{\perp \rm j} R_{\rm j} \right)}{J_m \left( k_{\perp \rm j} R_{\rm j} \right)} + i \pi \frac{m^2 / r_{\rm A j}^2}{\left| \rho  \Delta_{\rm A} \right|}} \nonumber \\
&& \times  \frac{J_m \left( k_{\perp \rm e} R_{\rm j} \right)}{H_m^{(1)} \left( k_{\perp \rm e} R_{\rm j}  \right)}, \label{eq:tmat}
\end{eqnarray}
where the prime $'$ denotes the derivative of the Bessel or Hankel function with respect to its argument.  In the absence of resonant damping, Equation~(\ref{eq:tmat}) consistently reverts to Equation~(17) of \citet{2009ApJ...692.1582L}.

On the other hand, the constants $\alpha_{2m}^{\rm j}$ satisfy an algebraic system of equations, namely 
\begin{equation}
\alpha_{2m}^{\rm j} + \sum_{\rm j' \ne j} \sum_{m' = - \infty}^\infty \alpha_{2m'}^{\rm j'} T_{m'm'}^{\rm j'} H_{m'}^{(1)} \left( k_{\perp \rm e} d_{\rm jj'} \right) \exp \left( i (m'-m) \varphi_{\rm jj'} \right) =0, \label{eq:system}
\end{equation}
for $-\infty < m < \infty$, where $\varphi_{\rm jj'}$ is the angle formed by the vector positions of the  centers of the two tubes with respect to the global reference frame. Equation~(\ref{eq:system}) is system with an infinite number of algebraic equations. The condition that there is a non-trivial solution, i.e., the determinant formed by the coefficients set equal to zero, provides us with the dispersion relation \citep[see details in][]{2009ApJ...692.1582L,2010ApJ...716.1371L}. The solution of the dispersion relation is the complex frequency of the damped collective quasi-mode. The imaginary part of the frequency is the resonant damping rate. In the case of \citet{2009ApJ...692.1582L,2010ApJ...716.1371L}, the solution frequency was real because of the absence of resonant damping. 

The dispersion relation is a  complicated expression and has to be solved by numerical methods. For practical computational purposes, in Equation~(\ref{eq:system})  the indices $m$ and $m'$ must be truncated into a finite number. The truncation term must be large enough to avoid inaccuracy of the solutions. Then, the dispersion relation can be solved using standard numerical routines to find the roots of transcendental equations.

\section{APPLICATION TO DAMPED OSCILLATIONS OF TWO LOOPS}
\label{sec:twoloops}

We recall that the T-matrix theory described in the previous Section is valid for an arbitrary number of loops. Here we  apply that method to investigate damped transverse oscillations in a loop system composed of two magnetic tubes alone. First we derive approximate expressions of the period and damping time in the case of two identical thin tubes. Later, we consider two tubes with different properties and arbitrary radii and perform a numerical study.

\subsection{Approximate solutions for two identical thin tubes}

We consider the paradigmatic case of two tubes with identical properties.   We denote by $\ri$, $R$, and $l$ the internal density, the radius, and the thickness of the boundary layer of both tubes, respectively. The external density remains $\re$. The distance between the centers of the two tubes is $d$.

 To obtain an approximate dispersion relation for kink-like modes we assume that the contribution from the azimuthal wavenumbers $m = \pm 1$ to the collective oscillation is much more important than the contributions from other values of $m$. Thus, we take $m,m' = \pm 1$  in Equation~(\ref{eq:system}) and roughly neglect the other terms. As pointed out by \citet{2009ApJ...692.1582L}, the coupling between the different values of $m$ gets stronger as the separation between the cylinders decreases. In terms of the parameters of the model, this means that our approximation applies to the case of large separations, i.e., $R/d \ll 1$. 

The system defined in Equation~(\ref{eq:system}) now becomes an algebraic system of four equations for the unknowns $\alpha_{2,-1}^{1}$, $\alpha_{2,1}^{1}$, $\alpha_{2,-1}^{2}$, and $\alpha_{2,1}^{2}$. We obtain the dispersion relation from  the condition that there is a non-trivial solution of the system, namely
\begin{eqnarray}
&& \left( H_0^{(1)} (k_{\perp \rm e} d)^2 -   H_2^{(1)} (k_{\perp \rm e} d)^2\right)^2 T_{11}^4 \nonumber \\
  &-& 2 \left( H_0^{(1)} (k_{\perp \rm e} d)^2 +   H_2^{(1)} (k_{\perp \rm e} d)^2\right) T_{11}^2 +1 =0, \label{eq:disper}
\end{eqnarray}
where we  used the property that $T_{11} = T_{-1 -1}$ according to the symmetry relations of the Bessel and Hankel functions \citep[see][]{abramowitz1972}. Equation~(\ref{eq:disper}) is an equation for $\omega$, which is enclosed in the definitions of $T_{11}$ and $k_{\perp \rm e}$. To go further analytically, we perform a series  expansion  of the Hankel functions for small arguments  and retain the first term alone. This would be approximately valid for thin tubes. After some algebraic manipulations, Equation~(\ref{eq:disper}) can be recast as
\begin{equation}
T_{11}  \pm i \frac{\pi}{4}(k_{\perp \rm e} d)^2 \approx 0. \label{eq:seriest11}
\end{equation}
Now, we look for a simplified expression of $T_{11}$. We use the TT approximation, i.e., $R/L \ll 1$. In Equation~(\ref{eq:tmat}) we perform an asymptotic expansion of the Bessel and Hankel functions  for small arguments. Equation~(\ref{eq:tmat}) becomes
\begin{eqnarray}
T_{mm} &\approx & - \frac{\frac{1}{\re \left( \omega^2 - k_z^2 \vae^2 \right)} - \frac{1}{\ri \left( \omega^2 - k_z^2 \vai^2 \right)}  + i \pi \frac{m/R}{\omega_{\rm R}^2\left| {\rm d} \rho / {\rm d}r \right|_R} }{\frac{1}{\re \left( \omega^2 - k_z^2 \vae^2 \right)} + \frac{1}{\ri \left( \omega^2 - k_z^2 \vai^2 \right)}  - i \pi \frac{m/R}{\omega_{\rm R}^2\left| {\rm d} \rho / {\rm d}r \right|_R}} \nonumber \\
&& \times \frac{i\pi}{2^{2m}} \frac{\left( k_{\perp \rm e} R \right)^{2m}}{ m! (m-1)!},  \label{eq:tmattt}
\end{eqnarray}
where we also assumed $r_{\rm A} \approx R$ for simplicity.  We take $m=1$, substitute Equation~(\ref{eq:tmattt}) into Equation~(\ref{eq:seriest11}), and arrive at the following expression,
\begin{eqnarray}
&& \ri  \left( \omega^2 - k_z^2 \vai^2 \right) \delta +  \re \left( \omega^2 - k_z^2 \vae^2 \right)  \nonumber \\
&-& i \frac{\pi}{R} \frac{\ri \re}{\left| {\rm d} \rho / {\rm d}r \right|_R} \frac{ \left( \omega^2 - k_z^2 \vai^2 \right) \left( \omega^2 - k_z^2 \vae^2 \right)}{\omega^2} = 0, \label{eq:reldisper}
\end{eqnarray}
where the parameter $\delta$ is defined as
\begin{equation}
\delta = \frac{1 \mp \left( \frac{R}{d} \right)^2}{1 \pm \left( \frac{R}{d} \right)^2}. \label{eq:delta}
\end{equation}

Equation~(\ref{eq:reldisper}) is the dispersion relation for resonantly damped kink modes of two identical  tubes in the TT and TB approximations and for large separations. In the limit that the tubes are  far from each other, $R/d \to 0$ and so $\delta \to 1$. Then, Equation~(\ref{eq:reldisper}) consistently reverts to the dispersion relation of resonant kink modes in an isolated thin tube \citep[e.g.,][]{1992SoPh..138..233G}. 

\subsubsection{Frequency of the Oscillations}

First we neglect the presence of the nonuniform boundary layers and study undamped oscillations. We drop the last term on the left-hand side of Equation~(\ref{eq:reldisper})  and the dispersion relation simplifies to
\begin{equation}
\ri  \left( \omega^2 - k_z^2 \vai^2 \right) \delta +  \re \left( \omega^2 - k_z^2 \vae^2 \right) = 0.
\end{equation}
The exact solution is
\begin{equation}
\omega^2 = \frac{\omega^2_k}{1 \mp \frac{\zeta -1}{\zeta + 1} \left( \frac{R}{d} \right)^2},  \label{eq:solundamped}
\end{equation}
where $\zeta = \ri/\re$ is the density contrast and $\omega_k$ is the frequency of the kink mode in an individual thin tube \citep{1983SoPh...88..179E}, namely
\begin{equation}
\omega_k^2 = \frac{\ri \vai^2 + \re \vae^2}{\ri + \re} k_z^2.
\end{equation}

Equation~(\ref{eq:solundamped}) can be compared to Equation~(51) of  \citet{2008A&A...485..849V}. Both expressions agree if the parameter $E$ of \citet{2008A&A...485..849V} is approximated as $E=\exp[-2 {\rm arccosh}(\frac{d}{2R})]  \approx \left(\frac{R}{d}\right)^2$, which is valid for $R/d \ll 1$. In agreement with  \citet{2008A&A...485..849V}, in the TT approximation we obtain two kink-like solutions corresponding to the $+$ and $-$ signs in Equation~(\ref{eq:solundamped}).  As explained by \citet{2009ApJ...692.1582L}, there are actually four kink-like modes beyond the TT approximation. For arbitrary radii,  the low-frequency solution (that with the $+$ sign  in Equation~(\ref{eq:solundamped})) splits in the $S_x$ and $A_y$ modes of \citet{2009ApJ...692.1582L}, while the high-frequency solution (that with the $-$ sign  in Equation~(\ref{eq:solundamped})) becomes their $S_y$ and $A_x$ solutions. 

From Equation~(\ref{eq:solundamped}) we compute the period of the oscillations, $P=2\pi/\omega$, as
\begin{equation}
P = P_{k} \sqrt{1 \mp \frac{\zeta -1}{\zeta + 1} \left( \frac{R}{d} \right)^2}, \label{eq:period}
\end{equation}
with  $P_k=2\pi/\omega_k$ the period of the  kink mode of an isolated tube.  For large separations the tubes feel little interaction. Then, the periods of the two solutions consistently tend to that of the  kink mode of an isolated tube.

\subsubsection{Damping Rate}

Here we incorporate the effect of resonant damping and consider the full expression of the dispersion relation (Equation~(\ref{eq:reldisper})). To obtain an approximate expression of the damping rate, we substitute $\omega = \omega_{\rm R} + i \omega_{\rm I}$ in Equation~(\ref{eq:reldisper}) and assume weak damping, i.e., $|\omega_{\rm I}| \ll |\omega_{\rm R}|$. Then, we are allowed to neglect terms of order $\omega_{\rm I}^2$ and higher orders. After some algebraic maniputations (not given here for the sake of simplicity) we arrive at an expression for the ratio $\omega_{\rm I} / \omega_{\rm R}$, namely
\begin{equation}
\frac{\omega_{\rm I}}{\omega_{\rm R}} = \frac{\pi}{2} \frac{1}{R} \frac{\ri \re}{\ri \delta + \re} \frac{1}{\left| {\rm d} \rho / {\rm d}r \right|_R} \frac{\left( \omega_{\rm R}^2 - k_z^2 \vai^2 \right) \left( \omega_{\rm R}^2 - k_z^2 \vae^2 \right)}{\omega_{\rm R}^2}. \label{eq:ratio1}
\end{equation}
To simplify Equation~(\ref{eq:ratio1}) we consider a smooth monotonic profile for the density in the nonuniform boundaries so that we can express the derivative of the density profile  as
\begin{equation}
\left| \frac{{\rm d} \rho}{{\rm d}r} \right|_R = \frac{\pi^2}{4} \mathcal{F} \frac{\ri - \re}{l},
\end{equation}
with $\mathcal{F}$ a factor that depends on the form of the profile. For instance, $\mathcal{F} = 4 / \pi^2$ for a linear profile  and $\mathcal{F} = 2 / \pi$ for a sinusoidal profile. In addition, we approximate $\omega_{\rm R}$ by the value of the frequency in the undamped case (Equation~(\ref{eq:solundamped})) and substitute the expression of $\delta$ (Equation~(\ref{eq:delta})) to arrive at
\begin{equation}
\frac{\omega_{\rm I}}{\omega_{\rm R}} \approx - \frac{1}{2\pi} \frac{1}{\mathcal{F}} \frac{l}{R} \frac{\zeta - 1}{\zeta +1} \frac{  1 \pm \left( \frac{R}{d} \right)^2 }{1 \mp \frac{\zeta -1}{\zeta + 1} \left( \frac{R}{d} \right)^2} \left[ 1 - \left( \frac{R}{d} \right)^4 \right]. \label{eq:ratio2}
\end{equation}
For a linear density profile, Equation~(\ref{eq:ratio2}) becomes Equation~(86) of \citet{2011A&A...525A...4R} if the approximation $\exp[-2 {\rm arccosh}(\frac{d}{2R})]  \approx \left(\frac{R}{d}\right)^2$ valid for $R/d \ll 1$ is again performed in their expression.

Now, we define the damping time as $\td = 1/ |\omega_{\rm I}|$ and use Equation~(\ref{eq:ratio2}) to obtain the expression for the ratio $\tdp$. By keeping terms up to $(R/d)^2$, the expression for the damping ratio is
\begin{equation}
\frac{\td}{P}  \approx \left( \frac{\td}{P} \right)_{\rm k} \left[ 1 \mp \frac{2 \zeta}{\zeta + 1} \left( \frac{R}{d} \right)^2 \right], \label{eq:ratio3}
\end{equation}
where
\begin{equation}
\left( \frac{\td}{P} \right)_{\rm k} = \mathcal{F} \frac{R}{l} \frac{\zeta + 1}{\zeta - 1},
\end{equation}
is the damping ratio of the individual kink mode \citep[see, e.g.,][]{2002ApJ...577..475R,2002A&A...394L..39G}. As  for individual kink modes, the damping ratio of collective modes is inversely proportional to the thickness of the nonuniform layer. Equation~(\ref{eq:ratio3}) shows that the solution with the $+$ sign is more efficiently damped by resonant absorption than the solution with the $-$ sign. This difference in the damping rates of the two modes qualitatively agrees with the results of \citet{2011A&A...525A...4R}. 


\subsection{Numerical Study}

Here we perform a numerical study  beyond the approximate limits studied before. To do so, we consider the full dispersion relation obtained from Equation~(\ref{eq:system}) by using the general expression of the T-matrix elements (Equation~(\ref{eq:tmat})). The truncated dispersion relation is then numerically solved. Convergence tests of the results have been performed to make sure that the truncation value of the azimuthal series is large enough for the error in the solutions to be negligible. In short, we found that for kink-like modes the truncation value, namely $m_t$, has no important effect unless $m_t < 5$ and the two loops are next to each other ($d/R \approx 2$). When $m_t > 5$ and/or the loops are far from each other, we essentially obtain that the results are independent of $m_t$. We have used $m_t=30$ in all computations given here, which reduces the error and assures the excellent converge of the solutions even when $d/R \approx 2$.

 The numerical method follows a two-step procedure.  First, we solve the dispersion relation in the absence of nonuniform boundary layers. In that case, the solution is a real frequency that corresponds to an approximation to $\omega_{\rm R}$. This approximate value of $\omega_{\rm R}$ is used  to compute the resonance positions and the derivative of the density profile at the resonances. We assume sinusoidal density profiles in the nonuniform boundary layers.  Then, we use these parameters to solve  the complete dispersion relation, which now includes the effect of resonant damping. The frequency obtained from the second run is complex, so that it provides us with a more accurate value of $\omega_{\rm R}$ and also gives us the value of $\omega_{\rm I}$.

\subsubsection{Identical loops}

We initially study the case of two identical tubes. We set the Cartesian coordinates system so that the $xy$-plane is perpendicular to the axes of the loops. In that plane,  the centers of the two loops are located on the $x$-axis. We use the notation introduced by \citet{2008ApJ...676..717L} to denote the four kink-like modes present in a two-loop configuration. The modes are labeled as $S_x$, $A_x$, $S_y$, and $A_y$, where $S$ and $A$ denote symmetric or anti-symmetric motions of the two loops, respectively, and the subscripts $x$ and $y$ indicate the main direction of polarization of the oscillations in the coordinates system defined above. The eigenfunctions of these four modes in the case of loops without nonuniform boundary layers can be found in Figure~2 of \citet{2008ApJ...676..717L}.

\begin{figure*}[!htp]
\centering
\includegraphics[width=.95\columnwidth]{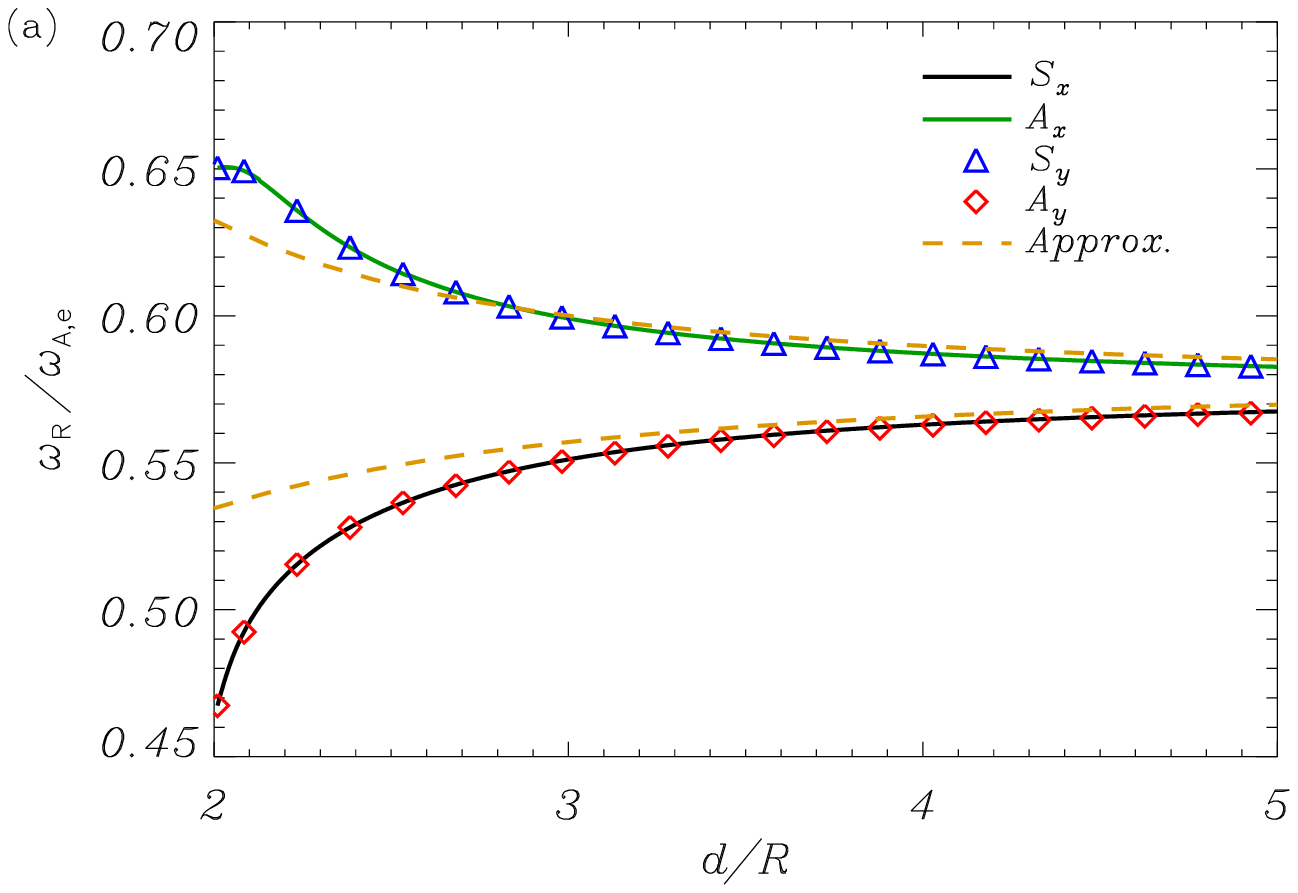}
\includegraphics[width=.95\columnwidth]{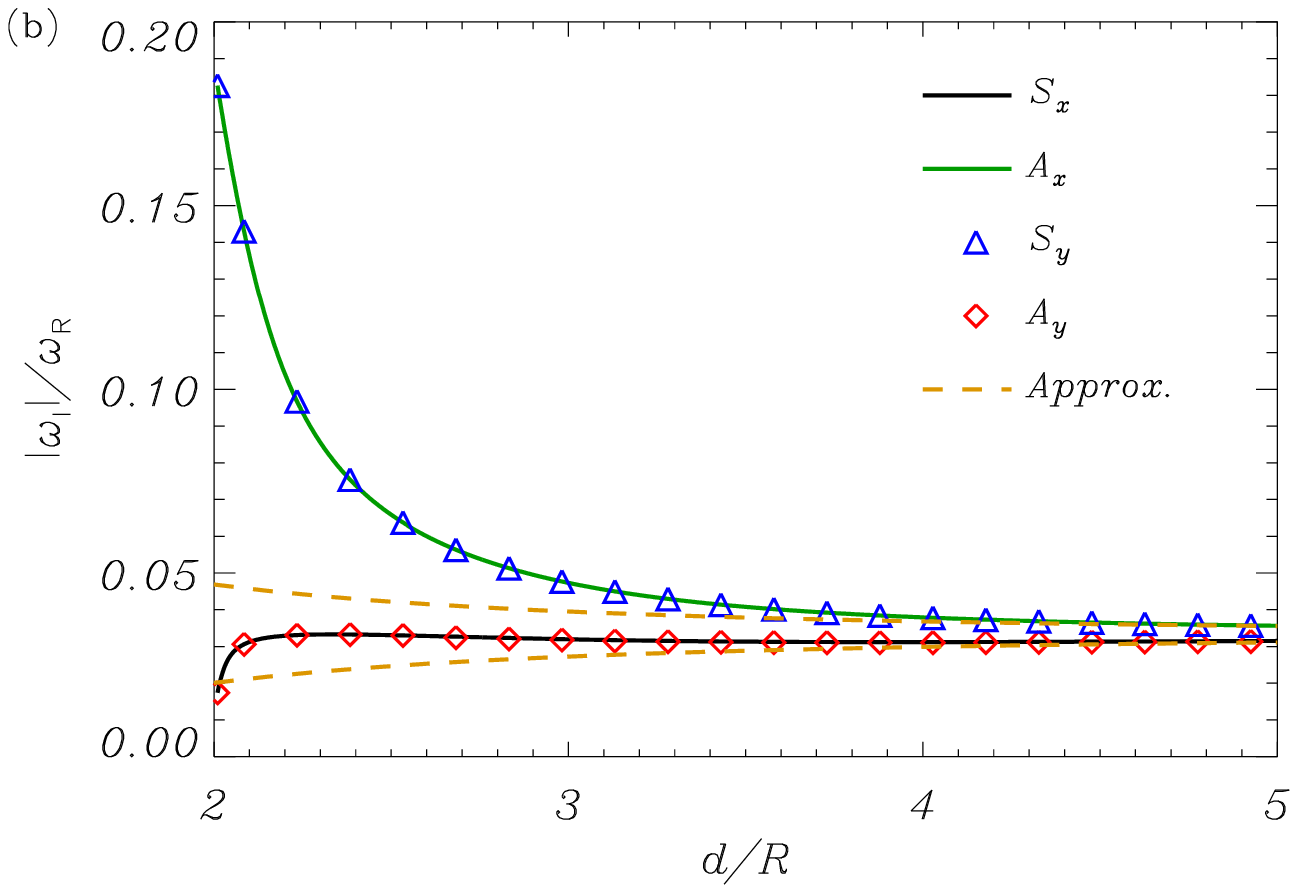}
\includegraphics[width=.95\columnwidth]{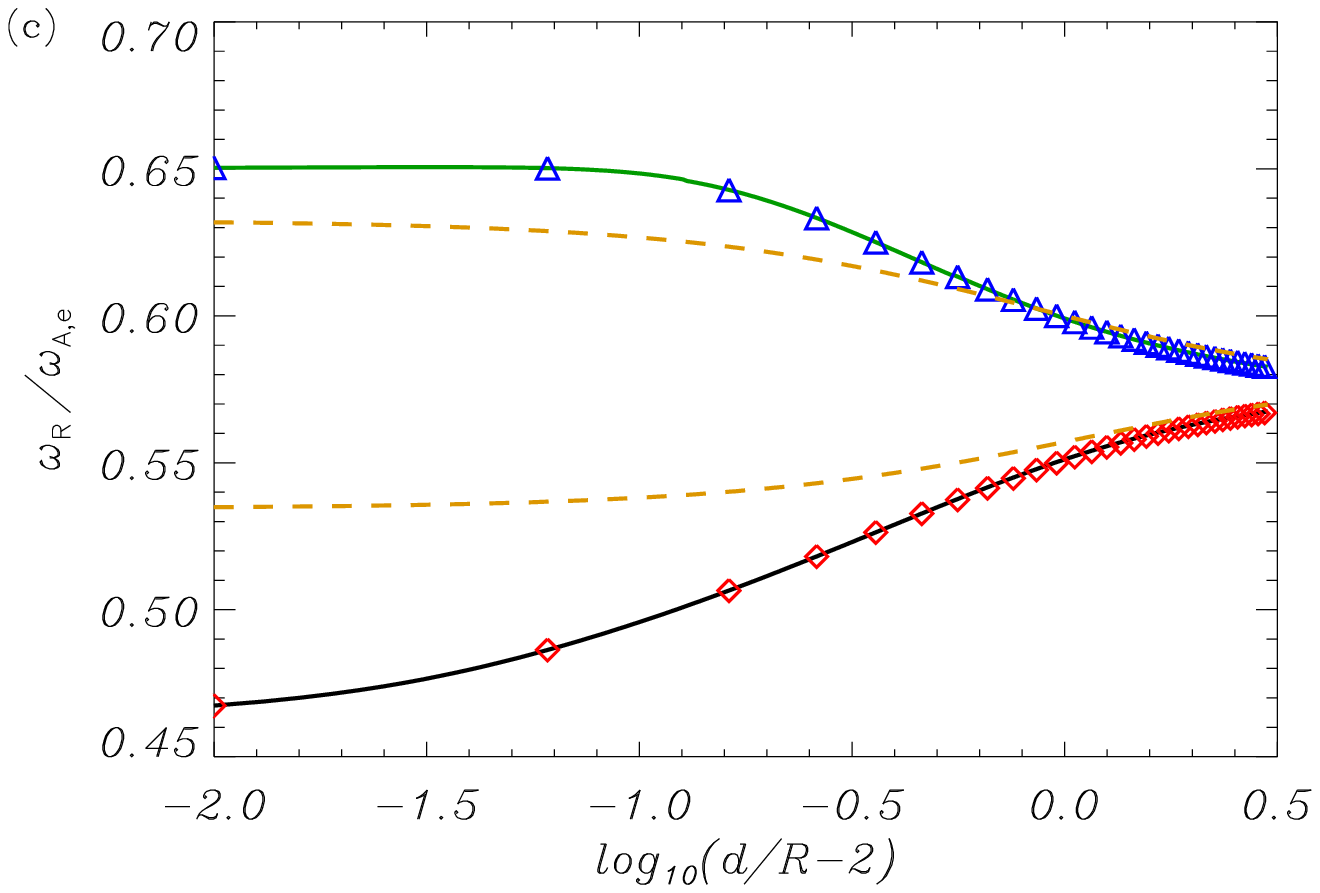}
\includegraphics[width=.95\columnwidth]{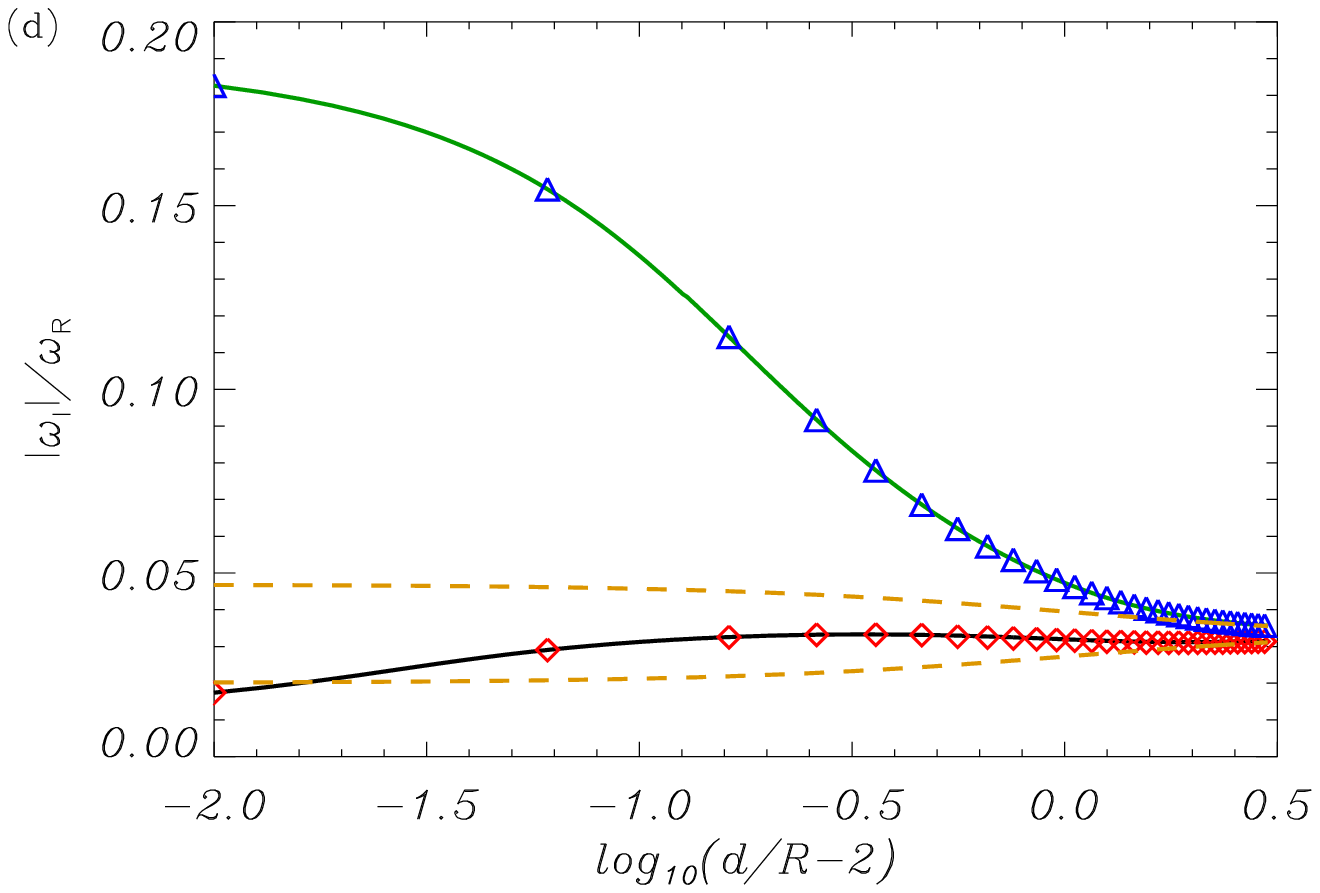}
\caption{Numerical results in the case of two identical coronal loops. (a) Dependence of $\omega_{\rm R}/\omega_{\rm A,e}$ on $d/R$, where $\omega_{\rm A,e} = k_z v_{\rm A,e}$ is the external Alfv\'en frequency. (b) Dependence of $\left|\omega_{\rm I}\right|/\omega_{\rm R}$ on $d/R$. The meaning of the various lines is indicated within the panels. The dashed lines correspond to the analytic approximations in the limit $d/R \gg 1$ (Equations~(\ref{eq:solundamped}) and (\ref{eq:ratio2})). We have used $\zeta = 5$, $L/R = 100$, and $l/R = 0.2$.   Panels (c) and (d) show the same results as panels (a) and (b), respectively, but as function of $\log_{10}\left(d/R-2\right)$. \label{fig:identical_d}}
\end{figure*}

\begin{figure*}[!htp]
\centering
\includegraphics[width=.95\columnwidth]{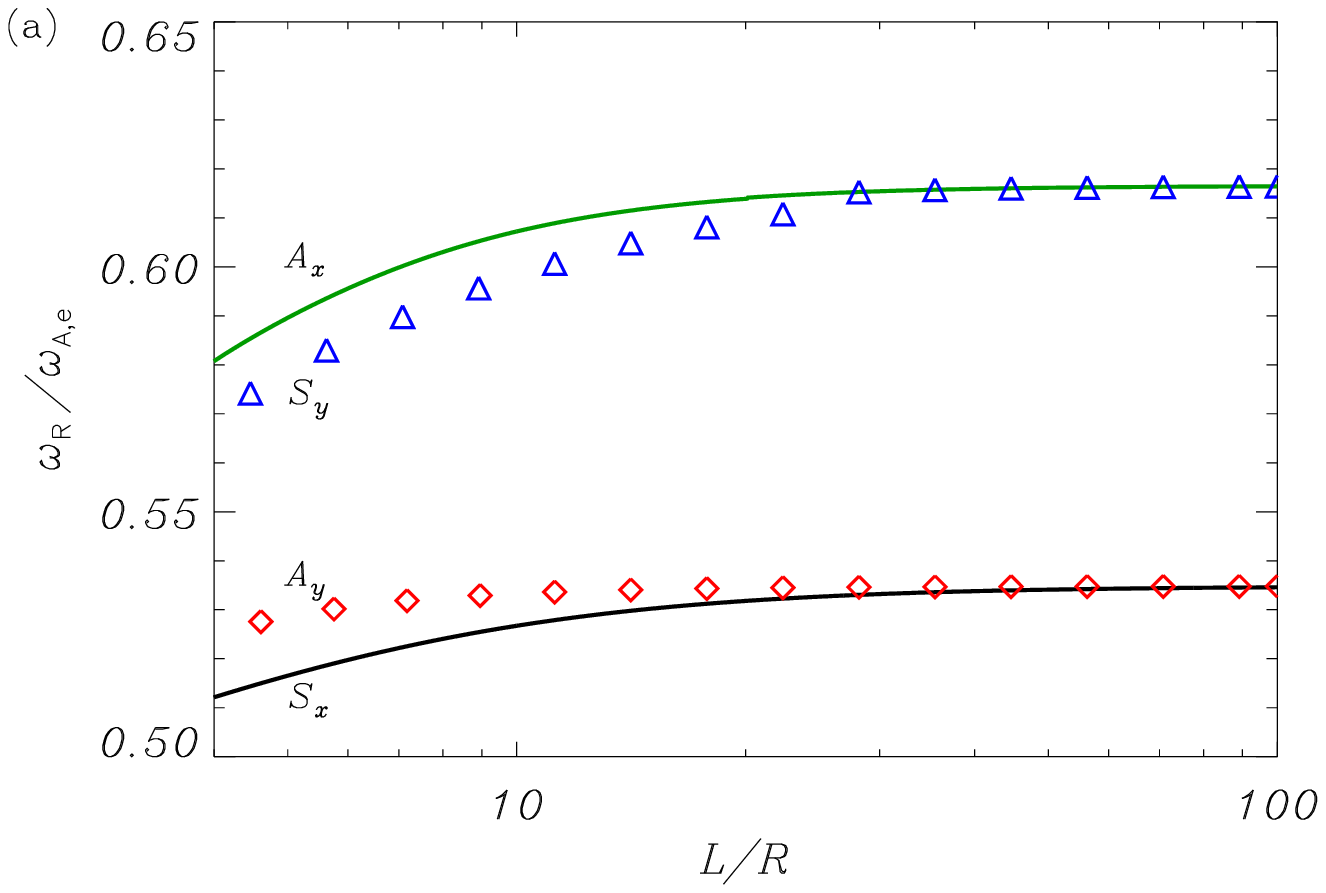}
\includegraphics[width=.95\columnwidth]{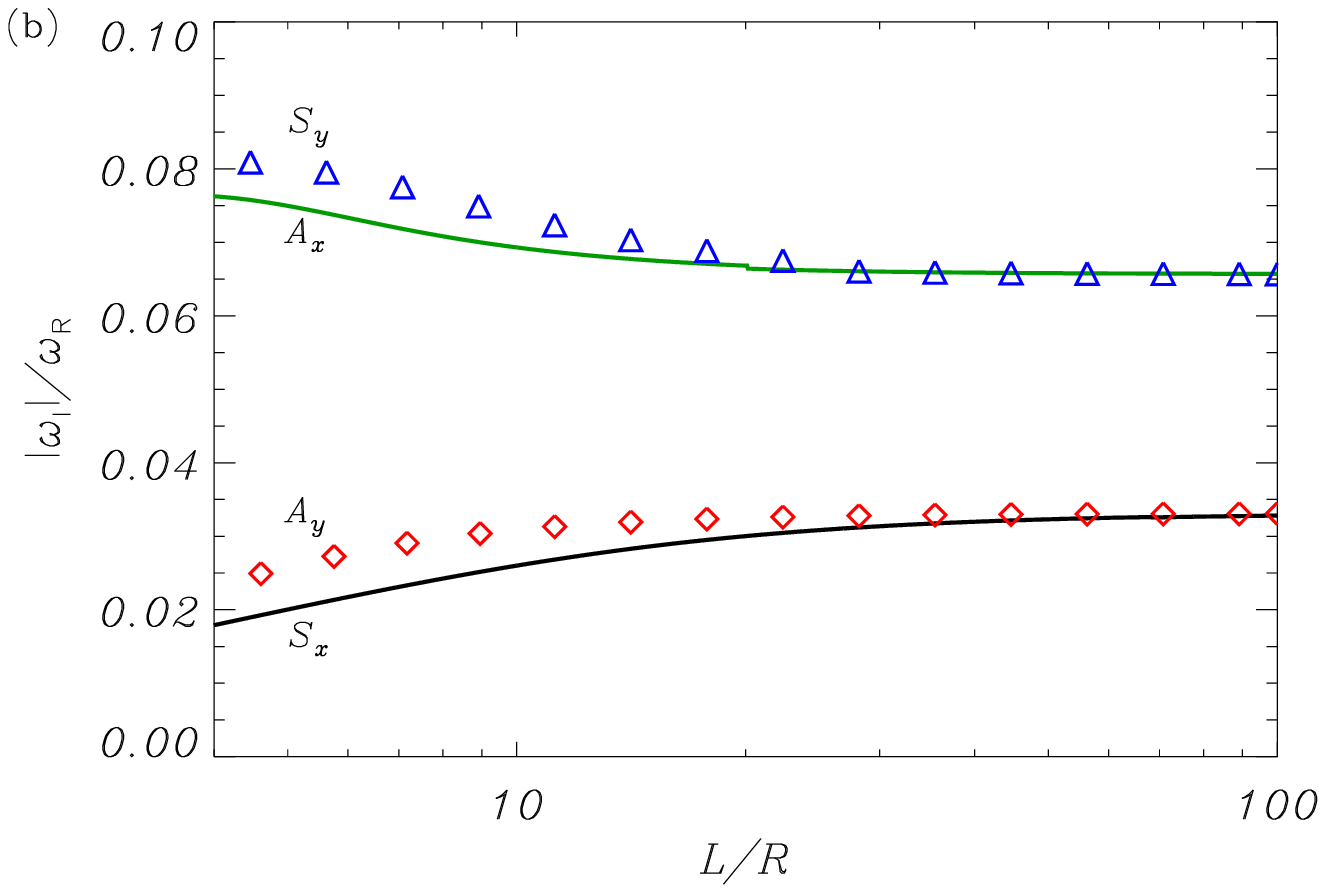}
\caption{Numerical results in the case of two identical coronal loops. (a) Dependence of $\omega_{\rm R}/\omega_{\rm A,e}$ on $L/R$, where $\omega_{\rm A,e} = k_z v_{\rm A,e}$ is the external Alfv\'en frequency. (b) Dependence of $\left|\omega_{\rm I}\right|/\omega_{\rm R}$ on $L/R$. The meaning of the various lines is indicated within the figure.  We have used $\zeta = 5$, $l/R = 0.2$, and $d/R = 2.5$. We note that the horizontal axes of both panels are in logarithmic scale.  \label{fig:identical_Len}}
\end{figure*}

Figure~\ref{fig:identical_d} shows the dependence of $\omega_{\rm R}$ and the ratio $\left|\omega_{\rm I}\right|/\omega_{\rm R}$ on the separation between loops, $d/R$, for a particular set of parameters given in the caption of the figure. We note that the curves corresponding to the $S_x$ and $A_y$ modes, and those of the $A_x$ and $S_y$ modes, are almost superimposed because we are in  the TT regime (we used $L/R = 100$). Concerning the behavior of $\omega_{\rm R}$,  the frequencies of the four solutions tend to the kink frequency of an isolated loop in the limit $d/R \gg 1$. Conversely, the smaller the separation between loops, the more important the splitting of the collective frequencies with respect to the kink frequency of an isolated loop. Figure~\ref{fig:identical_d}(a) can be compared to Figure~3 of \citet{2008ApJ...676..717L} and with Figure~4 of \citet{2008A&A...485..849V}. However, unlike in those previous works we note that in our case the frequencies of the high-frequency solutions ($A_x$ and $S_y$) do not tend to the external Alfv\'en frequency when $d/R \to 2$. The reason for this difference is probably that the $A_x$ and $S_y$ modes are strongly damped when $d/R \to 2$, and this fact has some impact on the real part of the frequency as well. 

On the other hand,  Figure~\ref{fig:identical_d}(b) shows that the high-frequency modes ($A_x$ and $S_y$) are more efficiently damped by resonant absorption than the low-frequency modes ($S_x$ and $A_y$). This result agrees with that of  \citet{2011A&A...525A...4R} for large separations. However, the behavior of  $\left|\omega_{\rm I}\right|/\omega_{\rm R}$ obtained here  for small separations is dramatically different from that of  \citet{2011A&A...525A...4R}. They found that the oscillations become undamped in the limit $d/R \to 2$, while we find that the modes remain damped. Although the analysis of  \citet{2011A&A...525A...4R} is mathematically correct, we find no physical reason for which the oscillations should be undamped when the loops are  close to each other. Our computations show that the damping  of the low-frequency modes ($S_x$ and $A_y$) is roughly independent on $d/R$, whereas the damping of the high-frequency modes ($A_x$ and $S_y$) gets stronger as the separation between loops is reduced. As pointed out by \citet{2011A&A...525A...4R} and \citet{2014A&A...562A..38G}, the physical significance of the results obtained with bycilindrical coordinates should be treated with  caution when the separation between the tubes is small.  

Panels (c) and (d) of Figure~\ref{fig:identical_d} show the same results as panels (a) and (b), respectively, but now $\log_{10}\left(l/R-2\right)$ is used in the horizontal axes. These additional graphs are included to  show in more detail the behavior of the solutions obtained with the T-matrix method for small separations between tubes.

We have overplotted in Figure~\ref{fig:identical_d} the analytic approximations of $\omega_{\rm R}$ and  $\left|\omega_{\rm I}\right|/\omega_{\rm R}$ given in Equations~(\ref{eq:solundamped}) and (\ref{eq:ratio2}). These approximations were derived in the limit $d/R \gg 1$ and  reasonably agree with the numerical solutions when $d/R \gtrsim 3$. As expected, the approximations do not work well for small separations. The analytic approximations were derived considering the contributions from $m=\pm 1$ alone, but  the contribution of high $m$'s to the full solution is important for small separation between loops \citep[see][]{2009ApJ...692.1582L}.

 Figure~\ref{fig:identical_Len} displays the dependence of the solutions on $L/R$.  This figure is included to show that the almost degenerate couples $S_x$--$A_y$ and $A_x$--$S_y$ split into four different solutions for small values of $L/R$ beyond the TT regime. We point out, however, that the impact of the value of $L/R$ on the solutions is not  relevant when realistic values of this parameter are considered. The TT limit used  by \citet{2011A&A...525A...4R} and \citet{2014A&A...562A..38G} is therefore adequate.

\begin{figure*}[!htp]
\centering
\includegraphics[width=.95\columnwidth]{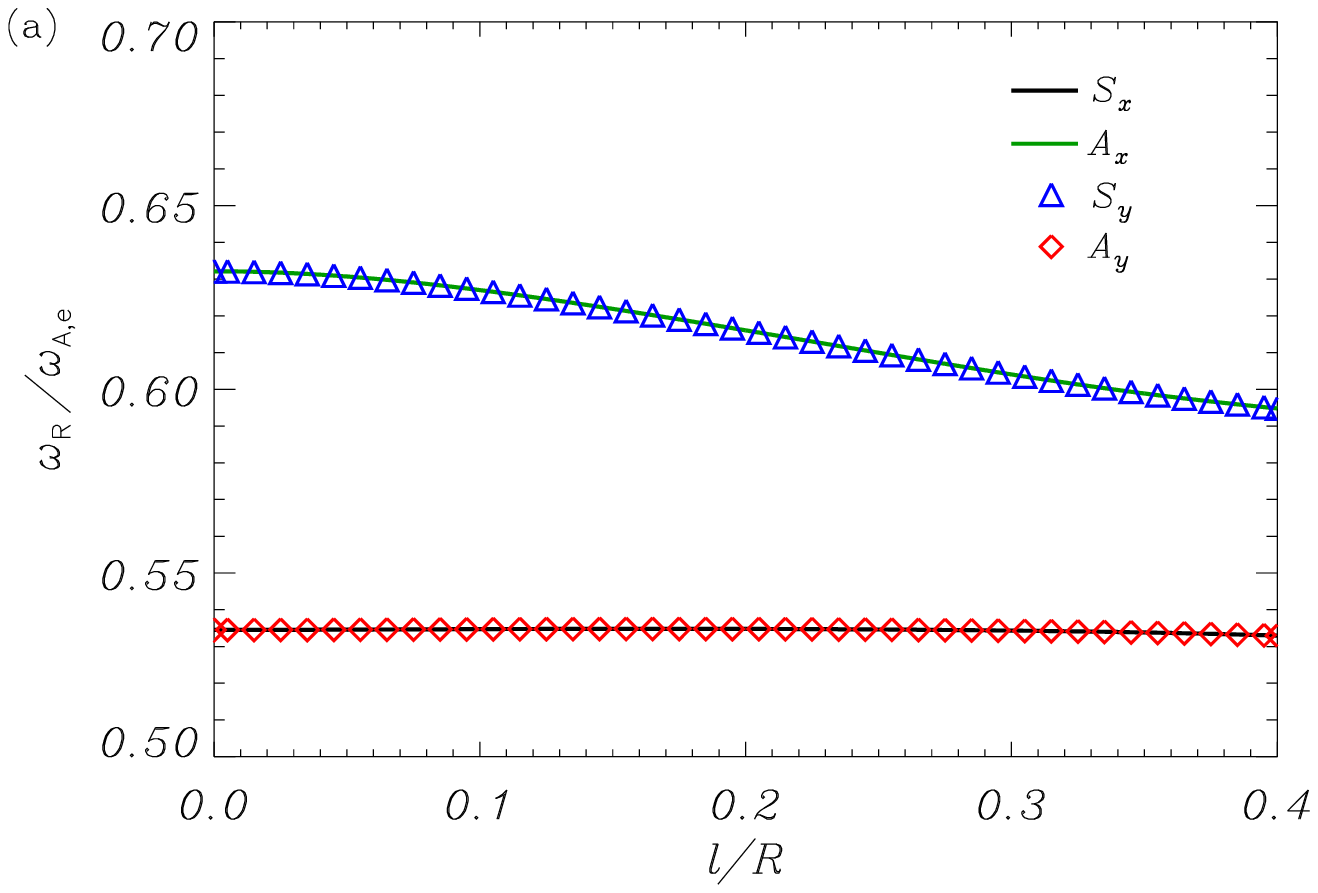}
\includegraphics[width=.95\columnwidth]{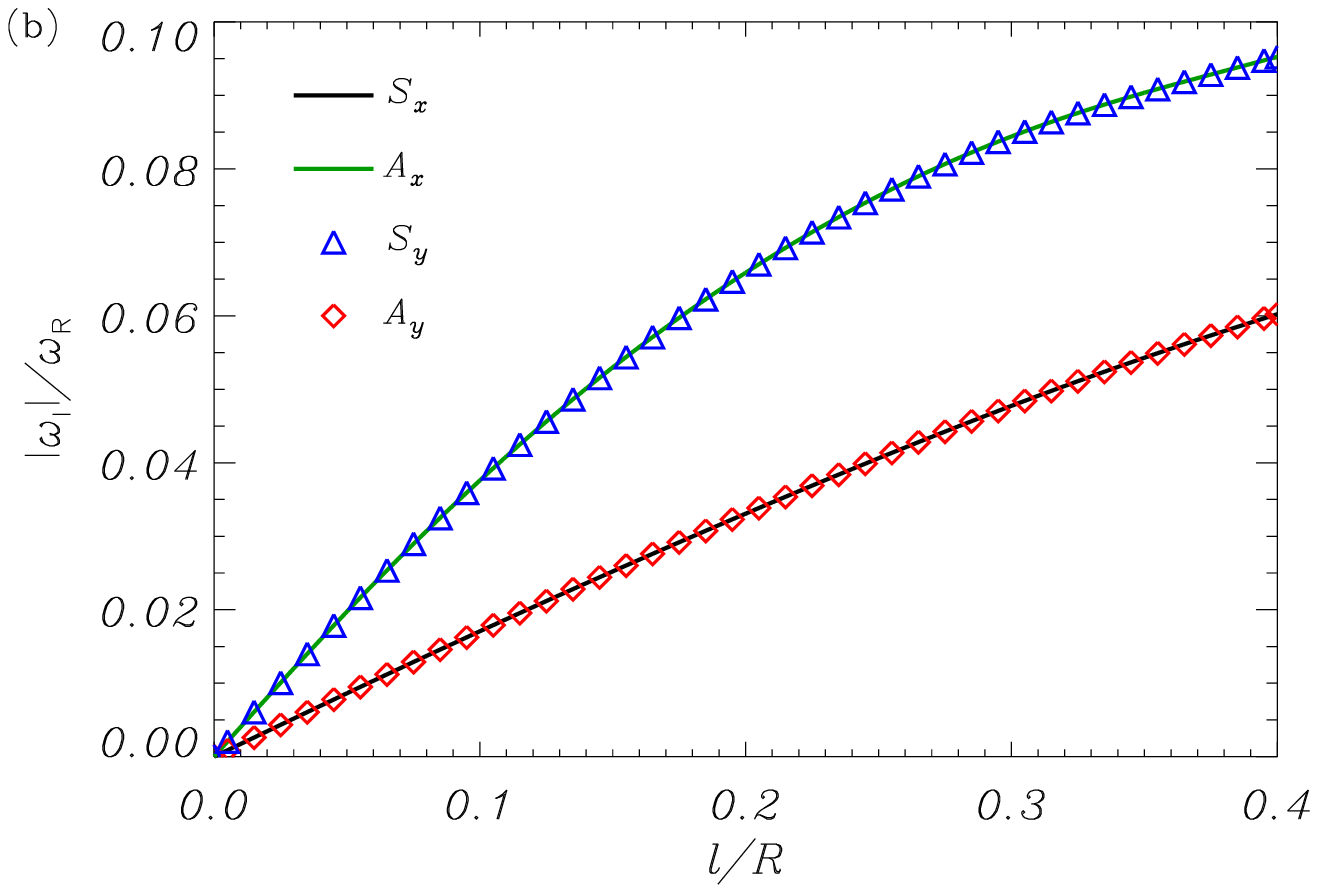}
\caption{Numerical results in the case of two identical coronal loops. (a) Dependence of $\omega_{\rm R}/\omega_{\rm A,e}$ on $l/R$, where $\omega_{\rm A,e} = k_z v_{\rm A,e}$ is the external Alfv\'en frequency. (b) Dependence of $\left|\omega_{\rm I}\right|/\omega_{\rm R}$ on $l/R$. The meaning of the various lines is indicated within the figure.  We have used $\zeta = 5$, $L/R = 100$, and $d/R = 2.5$.  \label{fig:identical_l}}
\end{figure*}

Now we plot in Figure~\ref{fig:identical_l} the dependence of $\omega_{\rm R}$ and $\left|\omega_{\rm I}\right|/\omega_{\rm R}$ on the nonuniform layer thickness, $l/R$. We consider a small separation, namely $d/R =2.5$, and the remaining parameters are the same as in Figure~\ref{fig:identical_d}.  Consistently, when $l/R=0$ the modes are undamped. The real part of the frequency of the low-frequency modes is almost independent of $l/R$, while their $\left|\omega_{\rm I}\right|/\omega_{\rm R}$ is roughly linear with $l/R$. Conversely, the real part of the frequency of the low-frequency modes decreases when $l/R$ increases, and their $\left|\omega_{\rm I}\right|/\omega_{\rm R}$ is only linear with $l/R$ for small values of this parameter. As discussed before, the behavior of the low-frequency modes ($S_x$ and $A_y$) is similar to that of the kink mode of an isolated loop. However, the high-frequency modes ($A_x$ and $S_y$) seem to be more affected by the interaction between loops and show a somewhat different behavior when $l/R$ increases. We note that because of the TB approximation  we are restricted to consider small values of $l/R$.

\begin{figure*}[!htp]
\centering
\includegraphics[width=.95\columnwidth]{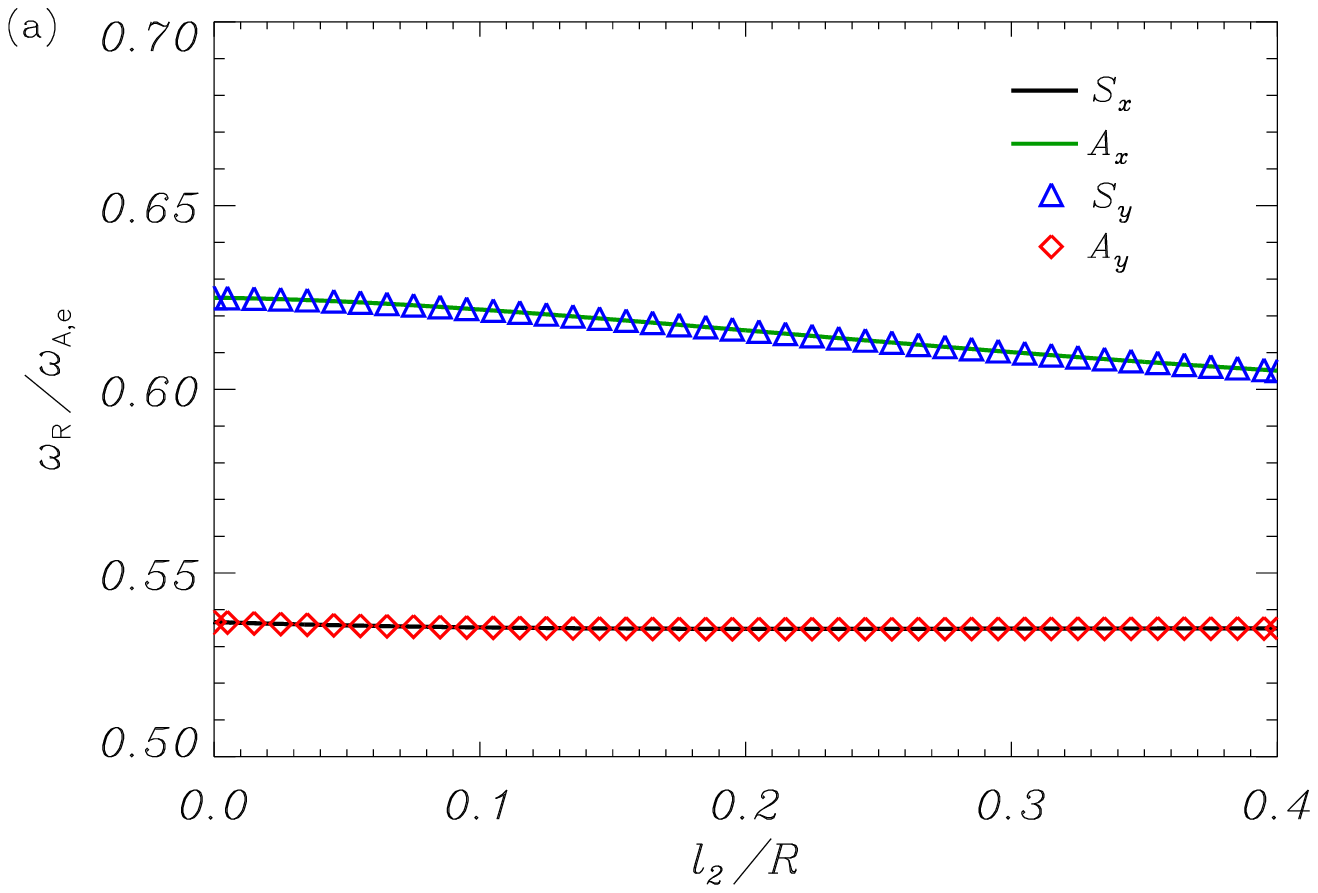}
\includegraphics[width=.95\columnwidth]{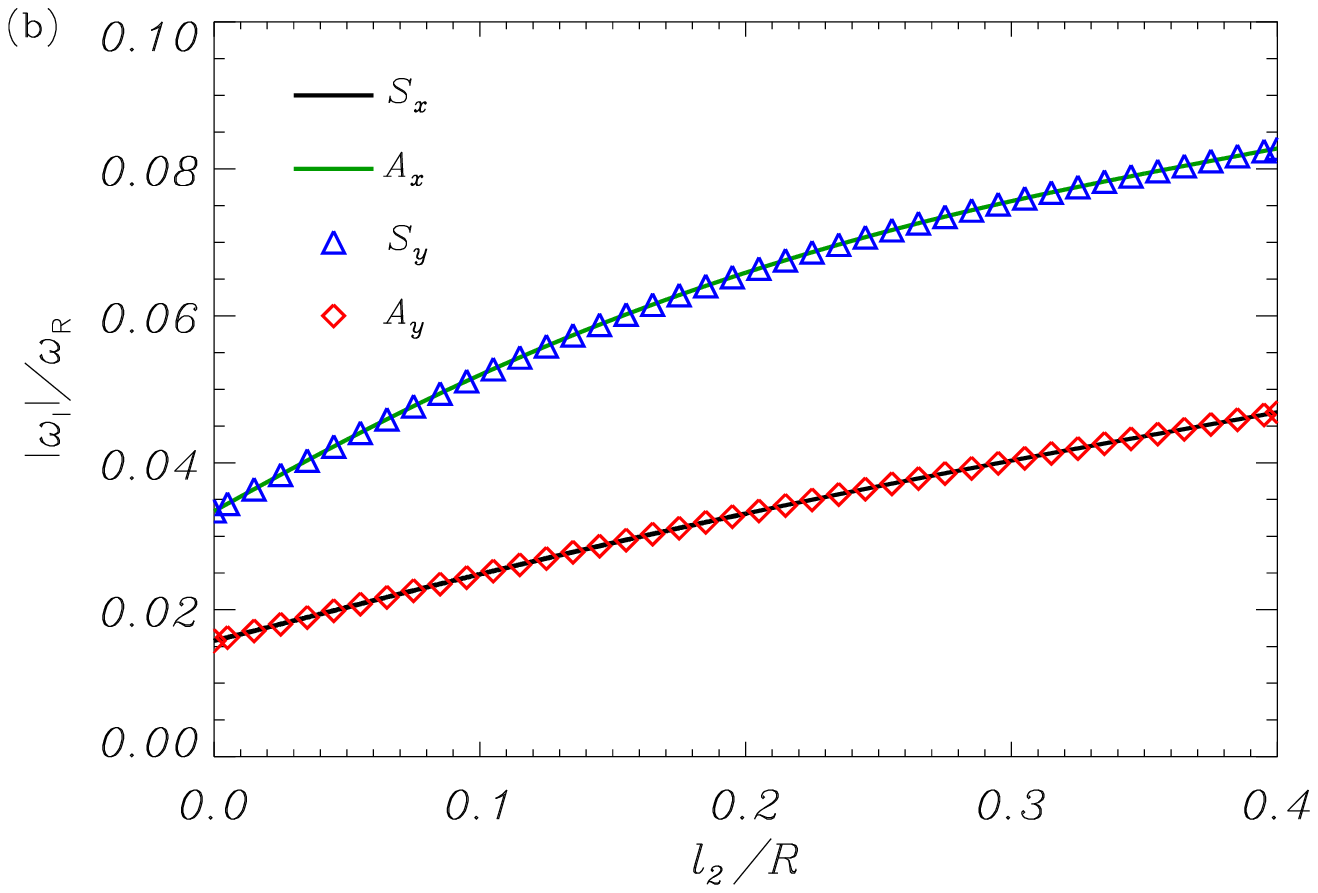}
\caption{Numerical results in the case of two non-identical coronal loops with $R_1=R_2=R$. (a) Dependence of $\omega_{\rm R}/\omega_{\rm A,e}$ on $l_2/R$, where $\omega_{\rm A,e} = k_z v_{\rm A,e}$ is the external Alfv\'en frequency. (b) Dependence of $\left|\omega_{\rm I}\right|/\omega_{\rm R}$ on $l_2/R$. The meaning of the various lines is indicated within the figure.  We have used $\zeta_1 = \zeta_2 = 5$, $L/R= 100$, $l_1/R = 0.2$ and $d/R = 2.5$.  \label{fig:nonidentical_l}}
\end{figure*}

It is useful to relate the present results with those of \citet{2007A&A...466.1145A,2008ApJ...674.1179A}, who studied the damping of  transverse oscillations of two nonuniform slabs. They found that the ratio $\left|\omega_{\rm I}\right|/\omega_{\rm R}$ corresponding to the symmetric kink mode of the two slabs is  weakly dependent of the separation between the two slabs \citep[see][their Figure~4]{2007A&A...466.1145A}. In turn, \citet{2008ApJ...674.1179A} found that the anti-symmetric kink mode of the two slabs is more efficiently damped than the  symmetric kink mode (see their Figure~6). The symmetric and anti-symmetric kink modes of  two slabs would be equivalent to the $S_x$ and $A_x$ modes of two cylinders. Thus, our results in cylindrical geometry are consistent with previous findings in Cartesian geometry. 

It is also convenient to consider the physical arguments of \citet{2007A&A...466.1145A,2008ApJ...674.1179A} to explain why the high-frequency modes damp more efficiently  than the low-frequency modes. \citet{2007A&A...466.1145A,2008ApJ...674.1179A}  related the efficiency of the damping with the magnitude of the total pressure perturbation within the resonant layers. According to \citet{2000ApJ...531..561A}, the efficiency of the resonant coupling between the global transverse mode and the Alfv\'en continuum modes is  proportional to the total pressure perturbation squared. \citet{2010PhDT.........3S} plotted the square of the total pressure perturbation corresponding to the $S_x$ and $A_x$ modes (see his Figures~9.7 and 9.8) and found that, when the quantities are normalized, the perturbation of the $A_x$ mode reaches a larger value in the resonant layers than that of the $S_x$ solution. This result qualitatively explains why resonant damping is more efficient for the $A_x$ mode  than for the $S_x$ mode. Equivalently, a similar reasoning help us  understand the different attenuation of the $S_y$ and $A_y$ modes. Nevertheless, a more robust study of the process of resonant absorption in two-dimensional configurations  would be needed for a  complete understanding of the different damping rates \citep[see][]{2010A&A...511A..17R}.

\subsubsection{Non-identical loops}

Here we consider two loops with different properties and compare our results to those of \citet{2014A&A...562A..38G}. We use subscripts 1 and 2 to refer to the two different loops.  For simplicity, we  take $R_1=R_2=R$ and $L/R=100$ in all following computations.

 First, we assume the same density contrast in the two loops, namely $\zeta_1 = \zeta_2 = 5$, and vary $l_2/R$ while $l_1/R$ is kept fixed to $l_1 / R = 0.2$. These results are shown in Figure~\ref{fig:nonidentical_l} and can be compared to those already displayed in Figure~\ref{fig:identical_l} in the case of $l_1 = l_2$. Importantly, we find that the collective oscillations remain damped when $l_2 / R = 0$. Although this is a rather particular situation, the results have interesting implications.  When $l_2 / R = 0$ resonant absorption only occurs in the boundary layer of   loop \#1. However, this is enough for the collective oscillations of the two loops to be efficiently damped. We note that in the study by \citet{2014A&A...562A..38G} the thicknesses of the nonuniform layers are linked to the coordinate system.

\begin{figure*}[!tp]
\centering
\includegraphics[width=.95\columnwidth]{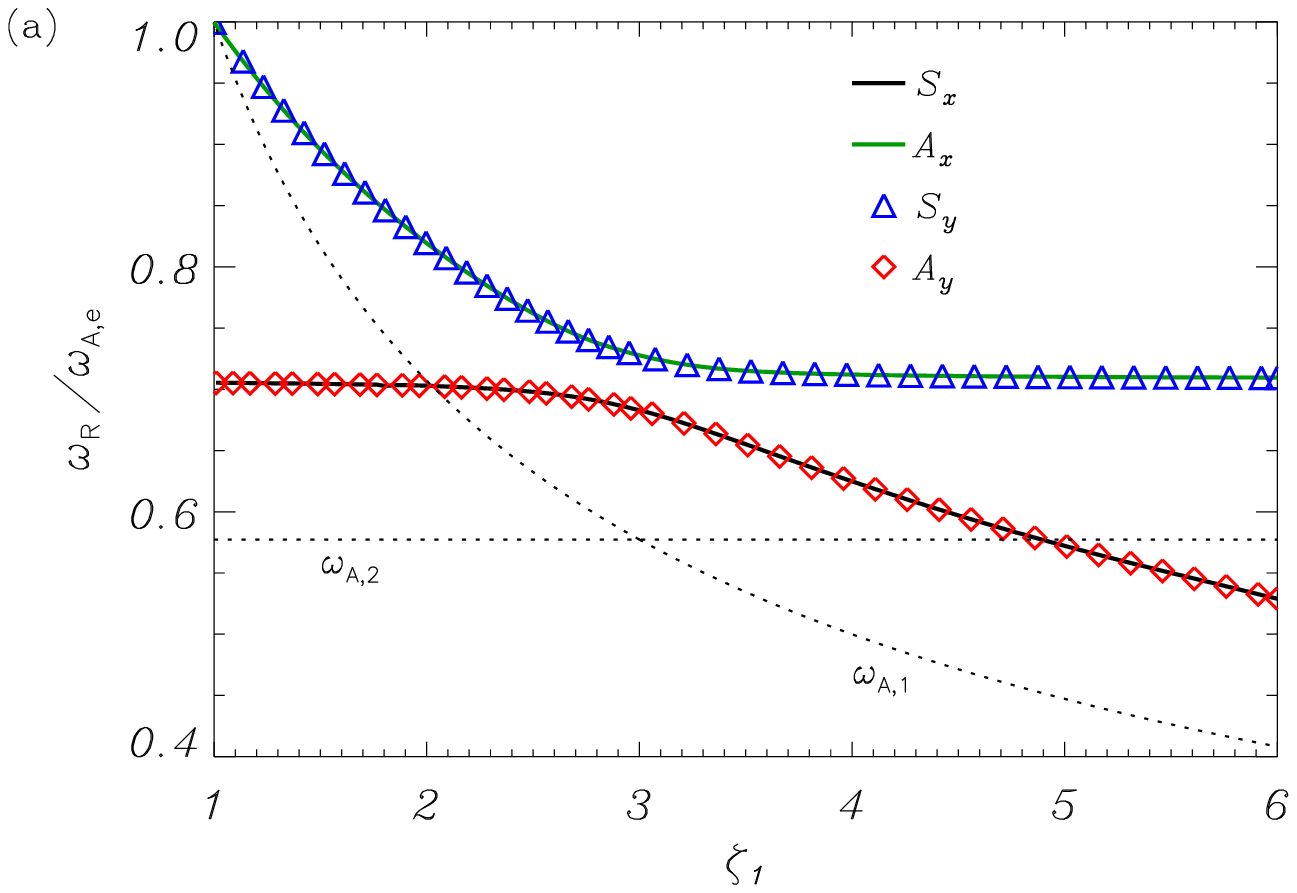}
\includegraphics[width=.95\columnwidth]{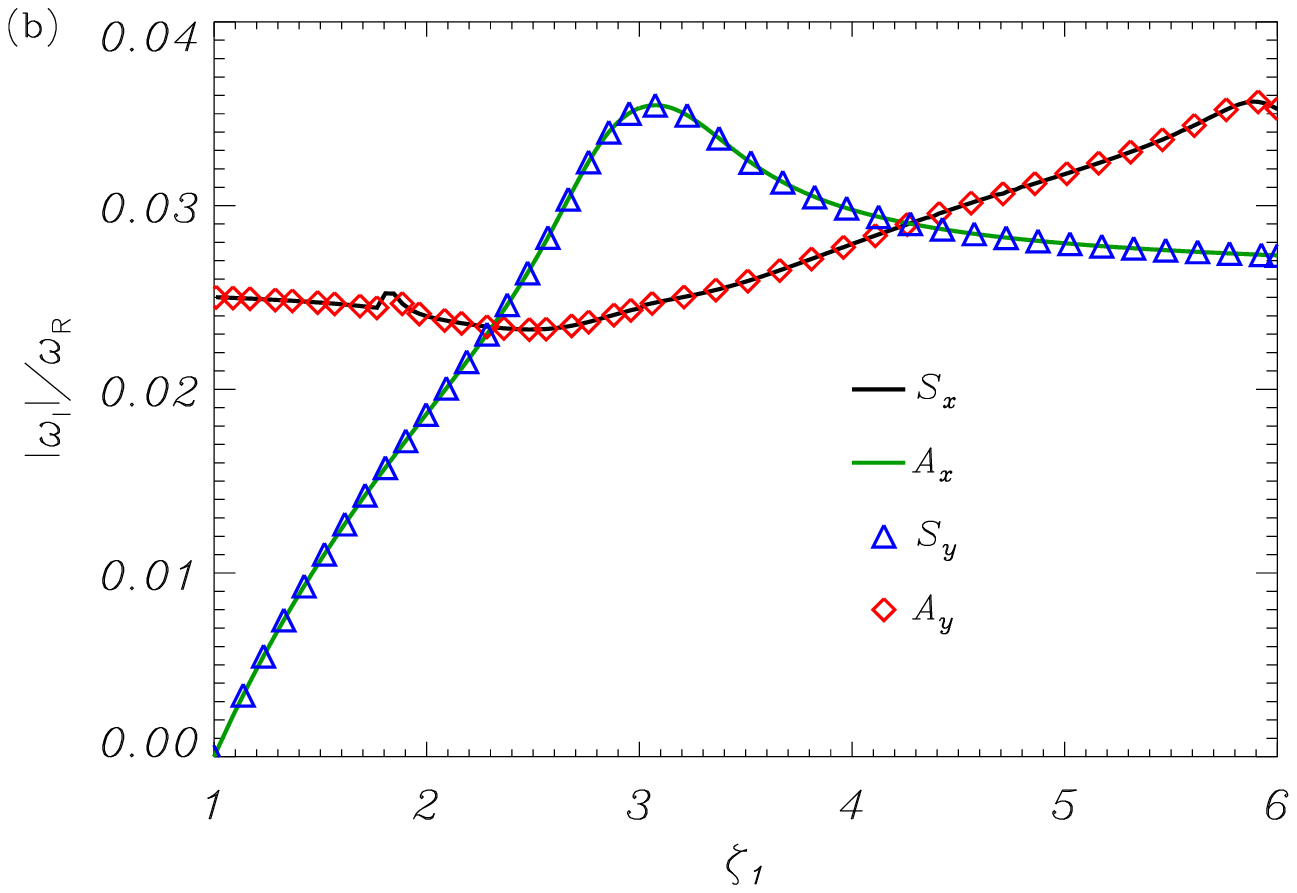}
\caption{Numerical results in the case of two non-identical coronal loops with $R_1=R_2=R$. (a) Dependence of $\omega_{\rm R}/\omega_{\rm A,e}$ on $\zeta_1$, where $\omega_{\rm A,e} = k_z v_{\rm A,e}$ is the external Alfv\'en frequency. (b) Dependence of $\left|\omega_{\rm I}\right|/\omega_{\rm R}$ on $\zeta_1$. The dotted lines correspond to the Alfv\'en frequencies of the two loops, and the meaning of the remaining lines is indicated within the figure.  We have used $\zeta_2 = 3$, $L/R = 100$, $l_1/R = l_2/R = 0.2$ and $d/R = 3$.  \label{fig:nonidentical_c}}
\end{figure*}

Now we consider the case of two loops with different density contrasts. We fix $\zeta_2 = 3$ and compute the solutions as functions of $\zeta_1$. These results are displayed in Figure~\ref{fig:nonidentical_c}, where the remaining parameters used in the computations are specified in the caption. For consistency, we keep the same notation as before to denote the various modes according to the ordering of their frequencies, although they do not  represent truly collective oscillations if $\zeta_1 \neq \zeta_2$ \citep{2009ApJ...692.1582L}. Figure~\ref{fig:nonidentical_c} can be compared to Figure~6 of \citet{2014A&A...562A..38G}. To make a proper comparison, we note that \citet{2014A&A...562A..38G} plotted the signed damping ratio, while here we plot the absolute value.  

Concerning the real part of the frequency (Figure~\ref{fig:nonidentical_c}(a)), we  find the same results as  \citet{2009ApJ...692.1582L}.  When $\zeta_1 < \zeta_2$, the high-frequency modes are associated to loop \#1 alone, whereas the low-frequency modes represent individual oscillations of loop \#2. The opposite happens when $\zeta_1 > \zeta_2$. Conversely, when $\zeta_1 \approx \zeta_2$ the four modes approach and interact in the form of an `avoided crossing'. Only in that case the modes represent truly collective oscillations  \citep{2009ApJ...692.1582L}.

Figure~\ref{fig:nonidentical_c}(a) also shows that the high-frequency modes are always within the Alfv\'en continua of the two loops, i.e., the frequencies of the $A_x$ and $S_y$ modes are always larger than $\omega_{\rm A,1}$ and $\omega_{\rm A,2}$ and smaller than $\omega_{\rm A,e}$. This is true except in the limit $\zeta_1 \to 1$, where the frequencies of the $A_x$ and $S_y$ modes tend to $\omega_{\rm A,e}$. On the contrary, the low-frequency $S_x$ and $A_y$ modes are below the Alfv\'en continuum of loop \#1 when $\zeta_1 \lesssim 2$ and below the Alfv\'en continuum of loop \#2 when $\zeta_1 \gtrsim 5$. This result may have implications for the damping by resonant absorption.

Figure~\ref{fig:nonidentical_c}(b) displays the damping ratio of the modes as a function of $\zeta_1$. The result for the high-frequency modes can be understood as follows. When $\zeta_1 \to 1$ the frequencies of the high-frequency modes tend to $\omega_{\rm A,e}$ and, as a consequence, these modes become undamped in that limit.  When $1 < \zeta_1 < \zeta_2$, the damping ratio increases when $\zeta_1$ increases. When $1<\zeta_1 < \zeta_2$ the high-frequency modes represent individual oscillations of loop \#1. Then, the damping ratio reaches a maximum when $\zeta_1 \approx \zeta_2$.  When $\zeta_1 > \zeta_2$ the damping ratio  saturates to a constant value because the high-frequency modes represent now individual oscillations of loop \#2, and the value of $\zeta_2$ is  fixed in the computations. The behavior of the damping of  the high-frequency modes agrees with that plotted by \citet{2014A&A...562A..38G} in their Figure~6.

The overall behavior of the damping ratio of the low-frequency modes  displayed in Figure~\ref{fig:nonidentical_c}(b) also agrees with \citet{2014A&A...562A..38G}. The damping ratio of the low-frequency modes is roughly constant when $\zeta_1 < \zeta_2$ and increases when $\zeta_1 > \zeta_2$.  The fact that the  low-frequency modes are below the Alfv\'en continuum of loop \#1 when $\zeta_1 \lesssim 2$ and below the Alfv\'en continuum of loop \#2 when $\zeta_1 \gtrsim 5$ have no important impact on the damping. Again, these results can be understood by considering that  the low-frequency modes are associated to loop \#2 when $\zeta_1 < \zeta_2$, while they are associated to loop \#1 when $\zeta_1 > \zeta_2$.

The low-frequency modes computed here do not show the pronounced minimum of the damping rate seen in the solution plotted by \citet{2014A&A...562A..38G}  when $\zeta_1 \approx \zeta_2$. There are several effects that may explain this difference. The most obvious one is  the different geometry considered in \citet{2014A&A...562A..38G}  and here. Another possible explanation is that the density profile in the nonuniform layers used by \citet{2014A&A...562A..38G}  is different from that used here. A linear density profile is used in \citet{2014A&A...562A..38G}, so that the derivative of density at the resonance position is independent of the frequency of the mode. Here we use a sinusoidal profile and  take into account that the position of the resonance  and the value of the derivative of density at the resonance position are functions of the mode frequency.

In Figure~\ref{fig:nonidentical_c}(b), the damping rate of the low-frequency modes shows a small bump around $\zeta_1 \approx 2$, when the real part of the frequency approximately crosses the internal Alfv\'en frequency of loop \#1. The reason for this bump is that the the low-frequency modes intersect with and `avoid cross'  the fluting modes that cluster toward the internal Alfv\'en frequency. This bump is absent from Figure~6 of \citet{2014A&A...562A..38G} probably because   coupling between kink and fluting modes is not described in the TT approximation.

\section{CONCLUDING REMARKS}
\label{sec:conclusion}

In this paper we have extended the analytic T-matrix theory of scattering  of \citet{2009ApJ...692.1582L,2010ApJ...716.1371L} to investigate resonantly damped oscillations of an arbitrary configuration of parallel cylindrical coronal loops. After presenting the general theory, we have performed a specific application in the case of two loops.  This work is partially based on unpublished results included in \citet{2010PhDT.........3S}, where collective damped oscillations of prominence threads were studied.

We have compared our results to those of the papers by \citet{2011A&A...525A...4R} and \citet{2014A&A...562A..38G}. They investigated the damping of collective oscillations of two loops in the TT approximation and used a method based on bicylindrical coordinates. In general, the results  of \citet{2011A&A...525A...4R} and \citet{2014A&A...562A..38G} are  in  good agreement with the present results, specially when the separation between loops is large. However, when the separation between the loops is small, i.e., for separations of few radii, the  results of those previous works show important differences compared to the present findings.  For instance, \citet{2011A&A...525A...4R} and \citet{2014A&A...562A..38G} obtained that by decreasing the distance between loops, the efficiency of resonant damping is reduced. In their computations, both low- and high-frequency modes become undamped when the  loops are in contact. However, this result lacks of a physical explanation and contradicts previous findings in Cartesian geometry \citep{2007A&A...466.1145A,2008ApJ...674.1179A}. In our computations, we find that the damping of the high-frequency modes gets stronger by decreasing the separation between loops, while the damping of the low-frequency modes is roughly independent of the separation. Our solutions do not become undamped when the two tubes are in contact. Thus, the results obtained here are consistent with previous results by \citet{2007A&A...466.1145A,2008ApJ...674.1179A} of collective oscillations of two slabs.

Although the mathematical analysis of \citet{2011A&A...525A...4R} and \citet{2014A&A...562A..38G}  is flawless,   their results by may be affected by unavoidable geometrical problems related to the bicylindrical coordinates when the loops are  close to each other. In bicylindrical coordinates the shapes of nonuniform boundary layers are not symmetric and change when the separation  between tubes decreases. The nonuniform layers get thicker in the outer parts of the tubes and thinner  in the inner parts.  As already mentioned  by \citet{2011A&A...525A...4R} and \citet{2014A&A...562A..38G}, these geometrical limitations may lead to unphysical results for small separations. The T-matrix method used here is not constrained by the geometrical problems of the bicylindrical coordinates. Therefore, we may conclude that the results given here are more generally applicable than those of \citet{2011A&A...525A...4R} and \citet{2014A&A...562A..38G} when the loops are close to each other.

Because of the TB approximation we were restricted to consider small values of $l/R$, i.e., $l/R \ll 1$. The effect of thick nonuniform layers could be included into the T-matrix formalism with the method of Frobenius used by \citet{2013ApJ...777..158S} in the case of an isolated loop. This would substantially increase the mathematical complexity of the problem but, on the other hand, it would provide a more accurate description of the damping of largely nonuniform loops. It has been shown by \citet{2014ApJ...781..111S} that the error in the damping rate associated to the use of the TB approximation can be important when the loops are largely non-uniform. Apart from a numerical factor, the damping rate in the TB approximation is  independent of the specific density profile considered within the nonuniform boundary, but the density profile can have a more important impact when the non-uniform layers are thick. In addition, the real part of the frequency depends on $l/R$ beyond the limit $l/R \ll 1$. The effect of thick nonuniform boundaries on collective loop oscillations could be explored in the future.

The  method given here to compute resonantly damped collective oscillations  can have multiple applications in the future. For instance, damped oscillations of a coronal arcade could be studied by modeling the arcade as a long line of parallel loops. Another interesting application is the investigation of oscillations of loops formed by many strands. As shown by \citet{2010ApJ...716.1371L}, the global oscillation of the whole loop would be determined by the interaction of the oscillations of the individual strands. The resonant absorption process working in the individual strands would affect the damping on the global loop motion \citep[see][]{2008ApJ...679.1611T}. In principle, the presence of multiple resonances in the system may cause the transverse oscillations of a multi-stranded loop to damp more quickly than the oscillations of an equivalent monolithic loop. This idea could be confirmed using the T-matrix method.  Also, as shown by \citet{2009ApJ...693.1601S},  the effects of gas pressure and mass flow along the loops can  easily be included in the T-matrix formalism, thus extending the applicability of the method.

\begin{acknowledgements}
We thank Ram\'on Oliver for useful comments on a draft of this paper. R.S. acknowledges support from MINECO through a `Juan de la Cierva' grant and through projects AYA2011-22846 and AYA2014-54485-P, from MECD through project CEF11-0012,  from the `Vicerectorat d'Investigaci\'o i Postgrau' of the UIB, and from FEDER funds. M.L. acknowledges support by MINECO through projects AYA2011-24808 and AYA2014-55078-P. M.L. is also grateful to ERC-2011-StG 277829-SPIA.
\end{acknowledgements}

\bibliographystyle{aa.bst} 
\bibliography{refs}

\end{document}